\documentclass[fleqn,usenatbib]{mnras}

\usepackage{newtxtext,newtxmath}

\usepackage[T1]{fontenc}

\DeclareRobustCommand{\VAN}[3]{#2}
\let\VANthebibliography\thebibliography
\def\thebibliography{\DeclareRobustCommand{\VAN}[3]{##3}\VANthebibliography}

\usepackage{graphicx}	
\usepackage{amsmath}	
\usepackage{comment}

\newcommand{\kms}{\mbox{km~s$^{-1}$}}

\newcommand{\mh}{\mbox{$\text{H}_{2}$}}
\newcommand{\pa}{\mbox{Pa-$\alpha$}}
\newcommand{\ha}{\mbox{H$\alpha$}}
\newcommand{\Njets}{\mbox{${24}$}}
\newcommand{\Nysos}{\mbox{${15}$}}
\newcommand{\Nhh}{\mbox{${13}$}}
\newcommand{\newHH}{\mbox{${7}$}}

\usepackage{soul}

\title[Outflows in NGC~3324]{Deep diving off the ‘Cosmic Cliffs’: previously hidden outflows in NGC~3324 revealed by \emph{JWST}}

\author[M. Reiter et al.]{Megan Reiter,$^{1}$\thanks{E-mail: Megan.Reiter@rice.edu (MR)}
Jon A. Morse,$^{2,3}$
Nathan Smith,$^{4}$
Thomas J. Haworth,$^{5}$ 
Michael A. Kuhn,$^{6}$
\newauthor
and Pamela D. Klaassen$^{7}$
\\
$^{1}$ Department of Physics and Astronomy, Rice University, 6100 Main St - MS 108, Houston, TX 77005, USA \\
$^{2}$ BoldlyGo Institute, 31 W 34th St, Floor 7 Suite 7159, New York, NY 10001, USA\\
$^{3}$ Visiting Associate in Astronomy, Division of Physics, Mathematics, and Astronomy, California Institute of Technology, Pasadena, CA 91125, USA\\
$^{4}$ Department of Astronomy, University of Arizona, 933 N. Cherry St, Tucson, AZ 85721, USA \\
$^{5}$ Astronomy Unit, School of Physics and Astronomy, Queen Mary University of London, London E1 4NS, UK\\
$^{6}$ Department of Astronomy, California Institute of Technology, 1216 East California Boulevard, Pasadena, CA 91125, USA \\
$^{7}$ UK Astronomy Technology Centre, Royal Observatory Edinburgh, Blackford Hill, Edinburgh EH9 3HJ, UK
}

\date{Accepted 2022 September 27. Received 2022 September 22; in original form 2022 September 5}

\pubyear{2022}

\begin{document}
\label{firstpage}
\pagerange{\pageref{firstpage}--\pageref{lastpage}}
\maketitle

\begin{abstract}
We present a detailed analysis of the protostellar outflow activity in the massive star-forming region NGC~3324, as revealed by new Early Release Observations (ERO) from the \emph{James Webb Space Telescope (JWST)}.  Emission from numerous outflows is revealed in narrow-band images of hydrogen Paschen-$\alpha$ (\pa) and molecular hydrogen. In particular, we report the discovery of \Njets \, previously unknown outflows based on their H$_2$ emission.
We find three candidate driving sources for these \mh\ flows in published catalogs of young stellar objects (YSOs) and we identify \Nysos\ IR point sources in the new \emph{JWST} images as potential driving protostars.  
We also identify several Herbig-Haro (HH) objects in \pa\ images from \emph{JWST}; most are confirmed as jets based on their proper motions measured in a comparison with previous \emph{Hubble Space Telescope} (\emph{HST}) \ha\ images.  
This confirmed all previous \emph{HST}-identified HH jets and candidate jets, and revealed \newHH\ new HH objects.
The unprecedented capabilities of \emph{JWST} allow the direct comparison of atomic and molecular outflow components at comparable angular resolution. 
Future observations will allow  quantitative analysis of the excitation, mass-loss rates, and velocities of these new flows. 
As a relatively modest region of massive star formation (larger than Orion but smaller than starburst clusters), NGC~3324 offers a preview of what star formation studies with \emph{JWST} may provide. 
\end{abstract}




\section{Introduction}

Jets and outflows are signposts of active star formation,  although details about their launching, quantitative connection to accretion, and timescales are still poorly understood.  
Jets can extend to lengths exceeding a parsec, making them readily identifiable in large-scale images with sufficient angular resolution and one of the most spectacular signposts of active star formation. 
They are launched by an underlying process of disk accretion \citep[e.g.][]{2006A&A...453..785F} and imprint a fossil record of its variations on the sky \citep[e.g.,][]{ell14}. 
This can indirectly indicate ongoing disk accretion in more distant ($>1$\,kpc) high-mass star-forming regions. 
Low-mass stars in high-mass star-forming regions are prime targets to quantify the impact of external feedback and are of particular interest to understand the impact of feedback on planet formation 
(e.g., \citealt{odell1993,odell1994,bally1998_hst,bally1998_disks,henney1998,johnstone1998,henney1999,bal00,bally2006,eisner2006,eisner2008,mann2010,mann2014,eisner2016,eisner2018}; see \citealt{2022arXiv220611910W,reiter2022} for recent reviews).
To date, most observations of the millimeter dust emission from disks around low-mass stars at these distances target jet-driving sources \citep[e.g.,][]{mesa-delgado2016,cortes-rangel2020,reiter2020_tadpole}.

As jets and outflows propagate from embedded young stellar objects (YSOs), 
they inject energy and momentum into the surrounding interstellar medium. 
The energy and momentum they inject
is predicted to resupply turbulence in clouds \citep{2007ApJ...662..395N, 2011ApJ...740..107C, 2014ApJ...784...61O, 2018MNRAS.475.1023M}. Jets also appear to play an important role in regulating the final masses of stars. 
For example, the \textsc{starforge} simulations predict that outflows have a non-negligible impact on the initial mass function (IMF) even in models that include other feedback mechanisms such as winds, radiation, and supernovae \citep[][]{2021MNRAS.502.3646G, 2022arXiv220510413G}. 

In high-mass star-forming regions, feedback from the most massive cluster members -- winds, radiation and supernovae -- creates large ionized bubbles \citep[e.g.][]{1977ApJ...218..377W, 1978ppim.book.....S,  2015NewAR..68....1D}. 
Once these other feedback mechanisms directly operate, jets/outflows make a much smaller contribution to the overall feedback budget. 
However, outflows act first, before the other feedback mechanisms initiate. 
At the periphery of feedback driven bubbles where molecular gas is still forming stars, outflows may dominate the local feedback. 
A full accounting of the different mechanisms and the timescales over which they dominate is important because feedback drives the evolution of the star-forming material \citep[e.g.][]{2010A&A...523A...6D, 2012MNRAS.427..625W, 2014MNRAS.442..694D, 2021MNRAS.501.4136A,  2022MNRAS.513.2088B}, may affect local star formation \citep[e.g.][]{2012MNRAS.421..408T, dale2015}, and enhances the distances out to which eventual supernovae impact the interstellar medium \citep[by carving low-density channels in the star-forming cloud, see e.g.,][]{2013MNRAS.431.1337R, 2020MNRAS.493.4700L, 2021MNRAS.504.1039R}.

Diffraction-limited data from the \emph{Hubble Space Telescope (HST)} has played a central role in the study of jets. 
Narrowband images reveal shock-excited emission from hydrogen recombination and forbidden emission lines in Herbig-Haro \citep[HH;][]{her50,her51,har52,har53} objects that trace shocks and jets especially once they have emerged from the neutral/molecular gas \citep[e.g.,][]{bur96,bal00,smith2010,rei13,rei14,reiter2016,reiter2017}. 
These images revealed detailed jet morphologies making clear the connection between disks and jets \cite[e.g.,][]{bur96}. 
In H~{\sc ii} regions, UV radiation from nearby high-mass stars illuminates the jet body, revealing un-shocked components and rendering the entire flow visible. 
Because of this, irradiated jets are a powerful tool for uncovering the mass-loss histories of the driving YSOs. 
Multi-epoch imaging traces measurable changes in the morphology and brightness distribution of shocks on human timescales and can be used to measure proper motions to determine (transverse) velocities \citep[e.g.,][]{har01,hartigan2005,hartigan2007,har11,hartigan2019}. 
Diffraction-limited imaging is key to identify collimated jets and outflows in more distant high-mass star-forming regions as these features are difficult to distinguish in seeing-limited images from the ground \citep[e.g.,][]{smith2004_hh666,smith2010}. 

With the arrival of the \emph{James Webb Space Telescope (JWST)}, similar studies are now possible for embedded jets and outflows seen only in the infrared (IR) that remain invisible at visual wavelengths. 
Near-IR observations reveal embedded portions of large-scale jets before they break out into the H~{\sc ii} region. 
Longer wavelengths also unlock the possibility of understanding the impact of jets/outflows in regions that will soon be overwhelmed with external feedback from nearby high-mass stars.

\emph{JWST} Early Release Observations \citep[ERO;][]{2022arXiv220713067P} targeted the massive star-forming region NGC~3324, providing an unprecedented look inside the star-forming gas surrounding the feedback-driven bubble.  
Only two confirmed and two candidate jets were previously identified toward this field in narrowband H$\alpha$ images obtained with \emph{HST} \citep{smith2010}. 
 Using the ERO images from \emph{JWST}, we have uncovered at least \Njets\ outflows in NGC~3324, primarily through their prominent emission in the F470N filter tracing shock-excited H$_2$. 
New F187N images of Paschen-$\alpha$ (\pa) emission trace many of the same features first seen in \ha\ images from \emph{HST}. 
The $\sim$16~yrs between the \emph{HST} and \emph{JWST} images allow these complementary observations to be used to provide the first proper motions of outflow features in this region. 
Together, this provides a first estimate of the kinematics of outflows in this portion of NGC~3324 and a unique look at embedded jets/outflows.

\section{Clarifying the relationship between NGC~3324 and Carina}\label{s:not_carina}
\begin{figure*}
	\includegraphics[width=\textwidth]{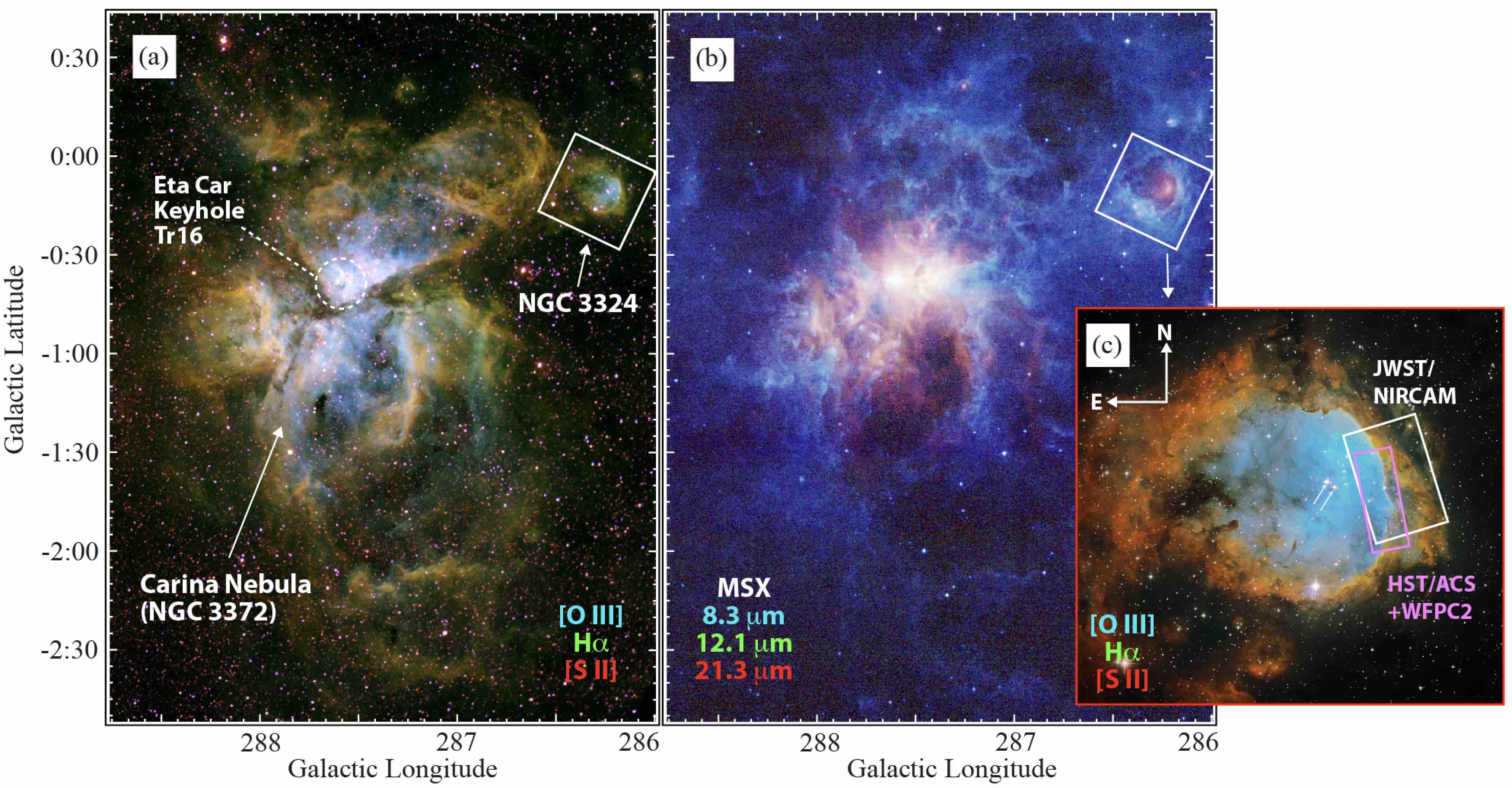}
    \caption{Images showing the location of NGC~3324 and the orientation of the JWST/NIRCam observations.  Panels (a) and (b) show very large scale overviews ($\sim$10 deg$^{2}$) of the Carina Nebula region as seen in visual-wavelength emission-line images and in mid-IR images from the Midcourse Space Experiment (MSX), respectively.  These panels are oriented in Galactic coordinates, and are adapted from figures in \citet{smith2000} and \citet{sb07}, and details of the images can be found there.  While the Carina Nebula is an extended Giant H~{\sc ii} region that dominates much of the field, NGC~3324 is the much smaller circular H~{\sc ii} region at right, and is a separate region outside the Carina Nebula (located about 1.5 deg or 60 pc from the center of Carina). The tilted white box in Panels (a) and (b) indicates the field of the more detailed image of NGC~3324 shown in panel (c), which is oriented along RA and DEC coordinates.  The image in Panel (c) uses the same colors/filters as Panel (a), obtained from Telescope Live\protect\footnotemark\ with permission (image credit: V.\ Unguru / Telescope Live).  The White box shows the JWST/NIRCam field of view, while the magenta box indicates the field of view of the HST images obtained with ACS+WFPC2. The two white arrows indicate two O-type stars that are the dominant sources of ionizing photons in NGC~3324 (see text). 
    }
    \label{fig:overviewBIG}
\end{figure*}

The NGC~3324 cluster and H~{\sc ii} region reside at the outskirts of the larger Carina star-forming complex (see Figure~\ref{fig:overviewBIG}).  The Carina Nebula has a well-established distance of about 2.3 kpc \citep{smith2006_distance,shull21}, and although it is outside the main nebula, NGC~3324 is thought to reside at a similar distance \citep{goeppl2022}.   

While the \emph{JWST} Early Release Observations were titled as images of the Carina Nebula, Figure~\ref{fig:overviewBIG} illustrates that NGC~3324 is actually not part of the Carina Nebula complex, and is instead a detached, circular H~{\sc ii} region \citep{smith2000}.  This distinction is important chiefly because the collection of more than 70 O-type stars that have powered feedback in the Carina Nebula -- including $\eta$ Carinae, 3 WNH stars \citep[Wolf-Rayet stars of the nitrogen sequence with H in their spectra, see][]{smith2008_WNH}, and several of the most luminous and  earliest O-types known \citep{smith2006_energy} -- are not the same stars that drive feedback in NGC~3324.  In Figure~\ref{fig:overviewBIG}b, red features (21 $\mu$m emission) locate hot dust grains in the interior of H~{\sc ii} region cavities that are heated by nearby O-type stars \citep{sb07}.  The hot dust in the interior of NGC~3324 is completely separate from the hot dust in Carina, and there does not appear to be a connection between the two cavities.  

The two sources that likely dominate ionization and feedback in NGC~3324 are HD~92206 and CPD$-$57$^{\circ}$3580, both of which are mid/late O-type multiple systems.  These two stars are indicated by white arrows in Figure~\ref{fig:overviewBIG}c.  SIMBAD\footnote{\href{https://simbad.u-strasbg.fr/simbad/}{https://simbad.u-strasbg.fr/simbad/}}
lists HD~92206 as an O6.5V+O6.5V binary, and CPD$-$57$^{\circ}$3580 as an O8V+O9.7V binary.  Note that the Galactic O-Star Catalog lists HD~92206 as five O-type stars contained in HD~92206A, B, and C \citep{sota14,ma16}, with HD~92206A resolved from HD~92206B (and HD~92206B is itself an O6+O6 binary), while the catalog also lists CPD$-$57$^{\circ}$3580 as HD~92206C (an O8+O8 binary).  In any case, feedback in NGC~3324 appears to be dominated by five closely spaced mid/late O-type stars in the center of the nebula, and it is not directly impacted by the high-mass stars in Carina.  

Overall, the combined ionizing photon luminosity of the stars that power NGC~3324 (2.8--3.5 $\times$ 10$^{49}$ ionizing photons per second; adopting the ionizing photon rates per spectral type in \citealt{smith2006_energy}) is about three times larger than the Orion Nebula, or about 3\% of the current ionizing photon luminosity of stars that power the Carina Nebula.

\section{Observations}
\subsection{\emph{JWST} ERO data}

\begin{figure*}
	\includegraphics[width=\textwidth]{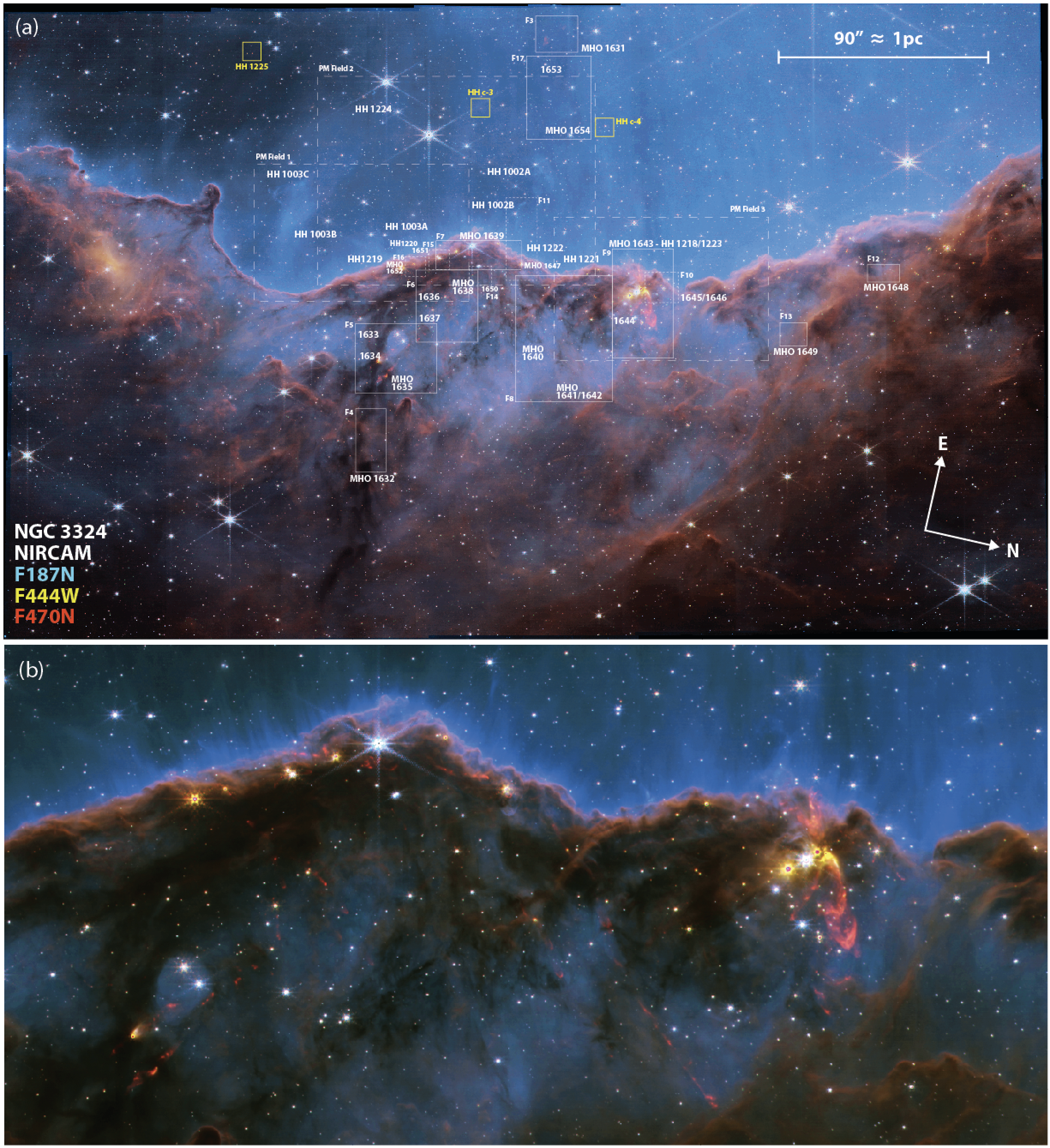}
    \caption{Overview image showing the location of various jets and molecular outflows in NGC~3324. The top panel (a) is the full field of view of the NIRCam image.  Color coding is F187 in blue, a combination of F187N and F444W in green (so that F187 appears teal and F444W appears yellow), and F470N in red.  Several boxes denote the locations of figures of individual outflows presented in this paper.  A label of "F2" indicates that this box corresponds to the enlarged region shown in Figure 2, etc.  Yellow boxes show locations of the three microjets. The bottom panel (b) shows an enlarged region of the central portion of the NIRCam image with no labels.  The color coding is the same as in (a), except that the red color is F470N with continuum suppressed somewhat (this is F470N flux with F444W scaled and subtracted to reduce the continuum emission to about 50\%, similar to continuum-subtracted images shown below but without fully subtracting the continuum). 
    }
    \label{fig:overview}
\end{figure*}
%

\begin{table}
	\centering
	\caption{\emph{JWST} observation log.}
	\label{t:obs}
	\begin{tabular}{lccrl} 
		\hline
		Instrument & Filter & Date & Exp.\ Time$^{\dagger}$ & Comment \\
		\hline
		NIRCam & F090W & 3 June 2022 & 3221.04 &   \\ 
		NIRCam & F187N & 3 June 2022 & 5797.86 &  Paschen-$\alpha$ \\ 
		NIRCam & F200W & 3 June 2022 & 3221.04 &   \\ 
		NIRCam & F335M & 3 June 2022 & 3221.04 &   \\ 
		NIRCam & F444W & 3 June 2022 & 3221.04 &   \\ 
		NIRCam & F470N & 3 June 2022 & 5797.86 & \mh\  \\ 
		MIRI & F1130W & 11 June 2022 & 6771.08 &   \\ 
		MIRI & F1280W & 11 June 2022 & 6993.12 &   \\
		MIRI & F1800W & 11 June 2022 & 5994.08 &   \\
		\hline
		\multicolumn{5}{l}{$^{\dagger}$ "effective exposure time" of the level 3 data products in seconds}
	\end{tabular}
\end{table}

In this paper, we use the ERO images obtained with \emph{JWST} as part of PID~2731 (PI: Pontoppidan). 
A summary of the observations, including the filters used and exposure times, is presented in Table~\ref{t:obs}.
For this study, we use the level 3 science products downloaded from the MAST archive (see Figure~\ref{fig:overview}). 
These data were produced with calibration software version 1.5.3 from the calibration context \texttt{jwst\_0916.pmap}. 
A more complete description of \emph{JWST} performance from commissioning may be found in \citet{rigby2022}.

Broad, medium, and narrowband images were obtained with the Near-Infrared Camera \citep[NIRCam;][]{2005SPIE.5904....1R, 2012SPIE.8442E..2NB} on 3 June 3 2022. 
The NIRCam mosaic covers a field $\sim 7.4\arcmin\ \times 4.4\arcmin$ on the edge of the NGC~3324 H~{\sc ii} region.  The location of this NIRCam field is identified by the white box in Figure~\ref{fig:overviewBIG}c, and a color composite of the full field of the NIRCam images (blue=F187N, green=F444W, red=F470N) is shown in Figure~\ref{fig:overview}a.
The 6.5~m diameter of \emph{JWST} provides an angular resolution of $\sim 0.07\arcsec - 0.17\arcsec$ over the range of NIRCam wavelengths considered in this study. 

Three broadband images of a subset of the area imaged by NIRCam were obtained with the Mid-Infrared Instrument \citep[MIRI;][]{2015PASP..127..584R,wright2015_miri} on 11 June 2022.  
The MIRI field is similar but smaller than  the NIRCam mosaics, covering an area 
$\sim 6.4\arcmin\ \times 2.2\arcmin$
that lies entirely within the NIRCam mosaic. 
The angular resolution of the MIRI images ranges from $\sim 0.44\arcsec - 0.70\arcsec$. 
We use these images for visual inspection of possible driving sources.

\footnotetext{\href{https://telescope.live/gallery/ngc3324-gabriela-mistral-nebula}{https://telescope.live/gallery/ngc3324-gabriela-mistral-nebula}}

\subsection{Archival \emph{HST} data}

The new images from \emph{JWST} cover and extend the area imaged in \emph{HST} PID~10475 during 7 and 21 March 2006 with the Advanced Camera for Surveys (ACS) in the F658N filter tracing H$\alpha$ + [N~II]. 
Full details of the ACS observations are presented in \citet{smith2010}. 
We use these archival H$\alpha$ images together with the new \pa\ images (F187N, see Table~\ref{t:obs}) from \emph{JWST} to measure proper motions of H-emitting knots and shock waves in the outflows (see Sec.~\ref{ss:jets}). The overlap between regions imaged with \emph{HST} and \emph{JWST} is shown in the right-most panel of Figure~\ref{fig:overviewBIG}.

\section{Results}\label{s:results} 

We report the discovery of \Njets\ H$_2$ flows in NGC~3324 based primarily on their emission in the F470N filter, and we identify or confirm several HH objects based on \pa\ emission in the F187N filter, or based on their motion when compared to previous \ha\ images obtained with \emph{HST}.  

Almost all of the H$_2$ flows are seen 
in the neutral/molecular clouds surrounding the H~{\sc ii} region cavity of NGC~3324; only 
three \mh\ features are seen in the H~{\sc ii} region.  New flows are cataloged as molecular hydrogen objects (MHOs)\footnote{\href{http://astro.kent.ac.uk/~df/MHCat/}{http://astro.kent.ac.uk/$\sim$df/MHCat/}}. 
We describe each new \mh\ flow in Section~\ref{ss:H2jets}. 

Several shock features are also seen in the H~{\sc ii} region, traced by \pa\ emission. 
Many of these features were previously identified in H$\alpha$ images from \citet{smith2010}.  
For features that are detected in both H$\alpha$ and \pa\ images, we use aligned frames to measure their proper motions (see Section~\ref{ss:jets}). 
Measuring proper motions also led to the detection of a few new features that, in retrospect, are also visible in the \ha\ images from \emph{HST}.

We use the \emph{Spitzer}/IRAC Candidate YSO Catalog for the Inner Galactic Midplane \citep[SPICY;][]{kuhn2021} to identify candidate driving sources. 
We also identify point sources on or near the flow axis in the new \emph{JWST} images and label them by their coordinates. 
A detailed analysis of their nature (including whether the spectral energy distributions; SEDs) are consistent with young sources will be presented in a future work. 
Candidate driving sources are listed in Tables~\ref{t:jets} and \ref{t:microjets}. 
The IR-bright stars that most likely drive the \mh\ outflows are readily identified in 17/\Njets\ cases. 
Two MHOs have ambiguous driving sources; 
four other features trace shock-like structures but do not have a clear origin or driving source.

\subsection{New H$_2$ flows}\label{ss:H2jets}

Newly flows presented in this section were discovered via their H$_2$ emission. 
Emission line features are especially prominent when the continuum or broadband emission from surrounding nebulosity is subtracted out \citep[see, e.g.,][]{reiter2016}. 
We use the wideband F444W filter to subtract continuum emission from the narrowband F470N filter. 
New MHOs are listed in Table~\ref{t:jets} and their locations are identified and labeled in the overview of the NIRCam image in Figure~\ref{fig:overview}a.
In the following, we briefly describe each of the new flows. 
\\

\textit{MHO~1631:}
%
\begin{figure}
	\includegraphics[width=\columnwidth]{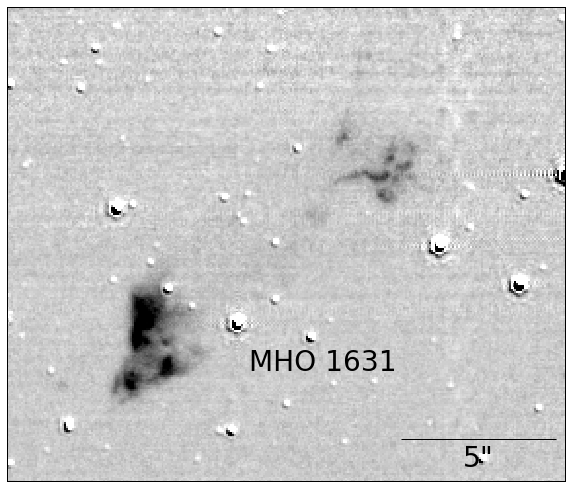}
    \caption{A continuum-subtracted H$_2$ (F470N--F444W) image of MHO~1631. The shock-like structure originates and extends $\sim$12.5\arcsec\ in the H~{\sc ii} region. No candidate YSOs are identified on or near the outflow axis.}
    \label{fig:MHOc1}
\end{figure}
An \mh-bright feature seen to be projected far out in the H~{\sc ii} region cavity is MHO~1631 (see Figure~\ref{fig:MHOc1}). 
It may trace a bow shock and its wake as it travels to the west (lower left in Figure~\ref{fig:MHOc1}). 
Alternately, it may be a flow 
propagating to the east (to the upper right in Figure~\ref{fig:MHOc1}), 
although there are no near-IR excess sources seen near MHO~1631 to confirm this orientation. 
The feature is seen close to the image edge, so additional knots may be outside the field of view. 
MHO~1631 is likely behind NGC~3324 as it does not have an ionized skin seen in \pa, as we would expect if it were in the H~{\sc ii} region. 
Neither MHO~1631 nor its host cloud are seen in extinction, as would be expected if it were a foreground object.
\\

\textit{MHO~1632:}
%
\begin{figure}
	\includegraphics[width=\columnwidth]{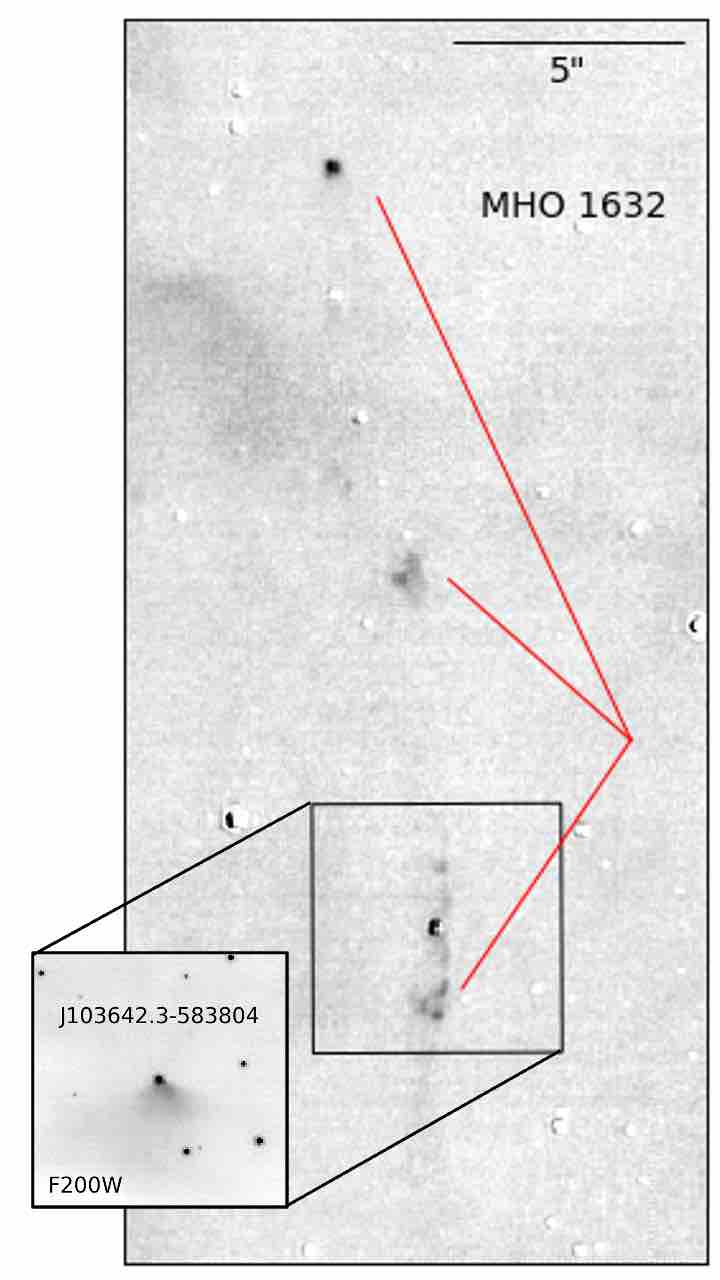}
    \caption{The MHO~1632 bipolar outflow that emerges from a point source, J103642.3-583804. Several shocks can be seen, tracing a slightly arced flow.  
    \textit{Inset:} Fan-like emission around the J103642.3-583804 seen in the F200W image.  
    }
    \label{fig:MHOc2}
\end{figure}
A few faint shock-like features trace the slightly arced MHO~1632 (see Figure~\ref{fig:MHOc2}). 
MHO~1632 represents the opposite extreme to MHO~1631 -- it is found at the deepest position in the cloud compared to the other MHOs (see Figure~\ref{fig:overview}). 
The flow is bipolar with emission on either side of the star on the outflow axis, J103642.3-583804. 
Diffuse, fan-like emission that opens from the driving source is seen in the F200W filter. 
This may trace reflected light in the outflow cavity if the western limb of the outflow (pointing down in Figure~\ref{fig:MHOc2}) is blueshifted.  
\\

\textit{MHO~1633:}
%
\begin{figure*}
	\includegraphics[width=\textwidth,trim=0mm 20mm 0mm 0mm]{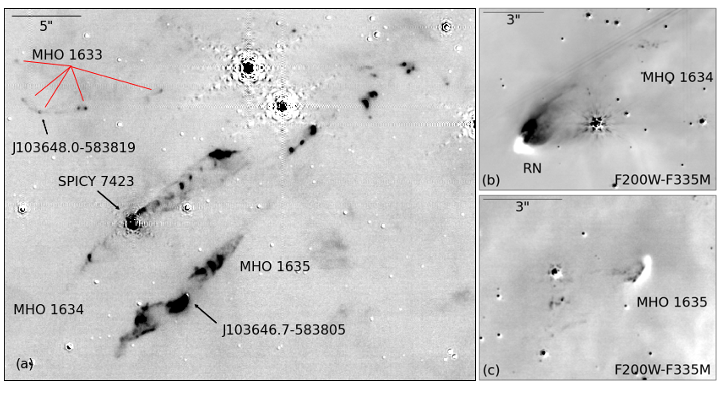}
    \caption{
    \textbf{(a)}
    Three outflows are seen in close proximity: MHO~1633, 1634, and 1635. 
    MHO~1633, seen in the upper left, is an asymmetric curved flow emerging from the point source J103648.0-583819. 
    Beneath it lies MHO~1634, a bipolar outflow emerging from SPICY~7423 and the spectacular reflection nebula that surrounds it. 
    \mh\ knots in MHO~1634 stand out prominently in the continuum-subtracted H$_2$ image. 
    Below MHO~1634 and with a nearly parallel outflow axis is MHO~1635. 
    Curved emission seen immediately below the candidate driving source J103646.7-583805 suggests the lower part of the outflow is blueshifted. 
    The upper portion of the outflow points toward another string of \mh\ knots that lies between MHO~1634 and MHO~1635, making it unclear to which flow they belong. 
    \textbf{(b)} F200W--F335M image of the reflection nebula at the origin of MHO~1634. Some knots can be seen in the upper right, likely due to shocked \mh\ in the F200W filter. 
    \textbf{(c)} F200W--F335M image of the origin of MHO~1635. Like MHO~1634, some knots are also visible. 
    }
    \label{fig:MHOc4}
\end{figure*}
A chain of \mh-bright knots trace the J-shape of MHO~1633 (see Figure~\ref{fig:MHOc4}).  
A YSO, J103648.0-583819 in the hook of the J-shape likely drives the flow. 
A bright knot of \pa\ emission at the location of the driving source may trace the base of the flow in the northern limb (right-hand side in Figure~\ref{fig:MHOc4}). 
The bend of the outflow is more dramatic than the other bent flows in this sample, more closely resembling examples seen in Orion \citep[e.g., LL~Ori and HH~336, see][]{bal01,bally2006}. 
Unlike the Orion jets, MHO~1633 is not directly exposed to strong winds and radiation in the H~{\sc ii} region. 
\\

\textit{MHO~1634:}
%
A bright reflection nebula associated with the YSO SPICY~7423 and nearby collimated \mh\ emission are immediately obvious in color images from \emph{JWST}. 
Continuum-subtracted H$_2$ images reveal a clear bipolar flow, MHO~1634, that bisects the reflection nebula (see Figure~\ref{fig:MHOc4}). 
The spacing of the inner knots appears quasi-periodic, similar to HD~163296 \citep{ell14}.
More diffuse emission surrounds these knots on either side, perhaps tracing the cavity walls of the outflow. 
Fainter arcs of \mh\ and a single bright knot trace the counterflow. 
An additional stream of fainter emission extends the counterjet to $\sim$6.5\arcsec\ (0.07~pc at a distance of 2.3~kpc) to the southwest of the driving source. 
\\

\textit{MHO~1635:}
%
Immediately below MHO~1634 in Figure~\ref{fig:MHOc4} lies another bright bipolar outflow, MHO~1635. 
A red, arcuate feature emerging from a dark cloud is clearly seen in color images. 
In the continuum-subtracted H$_2$ image, this arc appears to trace the base of a bipolar flow as it emerges from a deeply embedded YSO. 
A possible driving source, J103646.7-583805, is apparent in the MIRI data. 
Emission to the southwest (toward the bottom left of the image in Figure~\ref{fig:MHOc4}) is broad, as though tracing an outflow cavity. 
The red arc seen in color images traces one edge of this cavity. 
Continuous emission from J103646.7-583805 suggests that this southwest portion is the blueshifted side of the bipolar outflow. 
Additional emission extends to the southwest beyond this feature, extending $\sim$6.5\arcsec\ (0.07~pc) from the driving source. 

The MHO~1635 counterflow to the northeast (upper right in Figure~\ref{fig:MHOc4}) is also broad at first, but appears to get more collimated with distance from the driving source. 
The first knot of emission is offset $\sim$1\arcsec\ from the candidate driving source, suggesting that this is the redshifted limb propagating into the cloud. 
A series of bright \mh\ shock features delineate the flow axis before the inner flow terminates in fainter emission that tapers to a point. 
More collimation at large distances from the driving source is unusual, but has been observed in other sources \citep[e.g., HH~900,][]{smith2010,reiter2015_hh900}. 

The tapered tip of the inner flow points toward another chain of knots located $\sim$6.5\arcsec\ further to the northeast (upper right in Figure~\ref{fig:MHOc4}). 
These knots lie between the MHO~1634 and  MHO~1635 axes, making unclear from which flow they originate. 
This ambiguity may reflect the crossing of the MHO~1634 and MHO~1635 outflow axes.
\\

\textit{MHO~1636:}
%
\begin{figure}
    \includegraphics[width=\columnwidth]{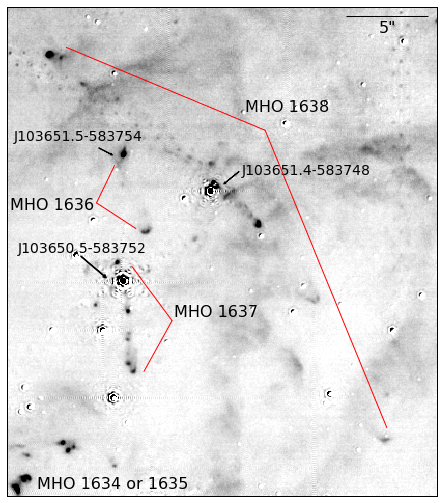}
    \caption{Another set of three outflows, MHO~1636, 1637, and 1638, lies to the northeast of the outflows in Figure~\ref{fig:MHOc4}. 
    Two knots, one bow-shock-shaped and one coincident with J103651.5-583754, constitute MHO~1636. 
    This flow lies above and traces a slightly different outflow axis than MHO~1637 below it. Several knots trace a collimated monopolar outflow axis beneath the source. A single knot above J103650.5-583752 may trace the counterflow or poor subtraction of the complicated \emph{JWST} PSF. 
    Arching above these two outflows is MHO~1638. The bipolar outflow emerges from J103651.4-583748 with multiple knots extending $>10\arcsec$ in either direction from the driving source.  
    Knots in the lower left corner of the image below to either MHO~1634 or MHO~1635. 
    }
    \label{fig:MHOc6}
\end{figure}
Another group of at least three separate outflows lies to the northeast of MHO~1634 and MHO~1635. 
The first, MHO~1636, consists of two bright H$_2$ knots (see Figure~\ref{fig:MHOc6}). 
One knot is coincident with the location of a point source, J103651.5-583754. 
The second has more of a bow-shock shape with faint wings that point back toward the YSO. 
No counterflow is apparent. 
While only two knots define MHO~1636, these knots lie away from the axes of the other outflows in the region. 
\\

\textit{MHO~1637:}
%
In Figure~\ref{fig:MHOc6}, MHO~1637 is seen emerging from a bright star, J103650.5-583752, located almost directly below MHO~1636. 
Several  H$_2$ knots trace a collimated outflow axis that extends to the west (down in Figure~\ref{fig:MHOc6}). 
Most of these knots have a tail of emission that points back in the direction of the driving source. 
A single knot on the opposite side of the YSO may trace the counterflow. 
However, this feature is located close to the star where artifacts from subtraction of the complex \emph{JWST} PSF may contaminate the emission.  
\\

\textit{MHO~1638:}
%
Several knots on either side of a point source, J103651.4-583748, trace the curved bipolar flow MHO~1638. 
Two point sources are seen near the flow axis in the NIRCam images; 
the fainter source lies on the axis. 
H$_2$ emission is bright and slightly extended at the location of the candidate driving source. 
More distant knots trace the bipolar flow as it extends to the northwest (to the upper left in Figure~\ref{fig:MHOc6}), pointing toward a bow shock $\sim$9\arcsec\ from the driving source. 
Complementary knots trace the southeastern lobe to a shock $\sim$12\arcsec\ from the driving source. 
Overall, MHO~1638 has a gentle C-shaped bend indicating deflection toward the west.
\\

\textit{MHO~1639:}
%
\begin{figure*}
	\includegraphics[width=\textwidth]{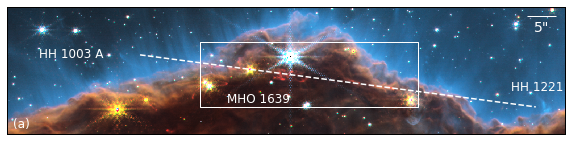}
	\includegraphics[width=\textwidth]{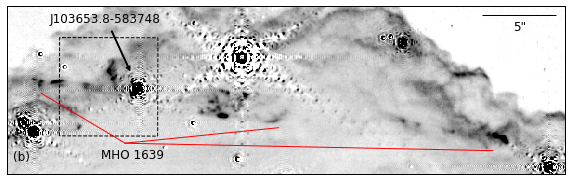}
	\includegraphics[width=0.325\textwidth]{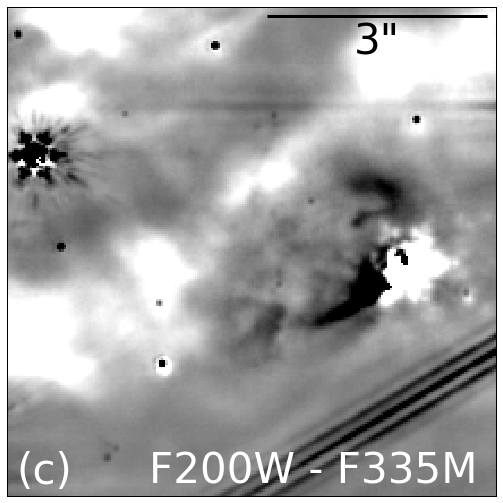}
	\includegraphics[width=0.325\textwidth]{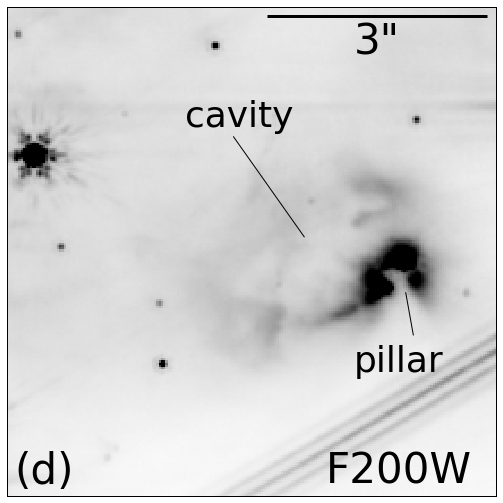}
	\includegraphics[width=0.325\textwidth]{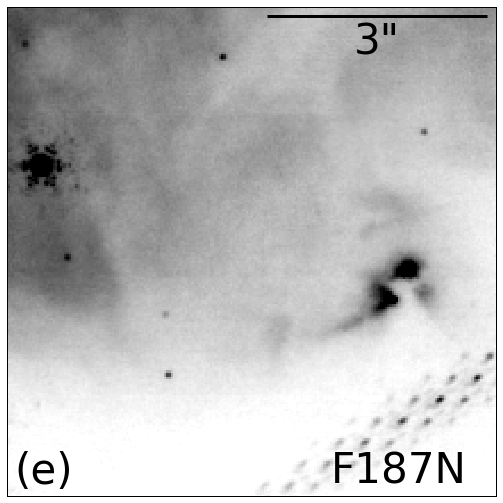}
    \caption{
    \textbf{(a)} Color image (same color mapping as in Figure~\ref{fig:overview}) showing \mh\ emission from MHO~1639 in the cloud (red) and the two bow shocks seen in \pa\ outside the cloud that appear to be part of the same flow, HH~1003~A and HH~1221 (blue). 
    A white dashed line connecting the apices of the HH~1003~A and HH~1221 bow shocks runs through the IR source J103653.8-583748 and bisects the \mh\ flow MHO~1639. A white box indicates the area shown in panel (b). 
    MHO~1651, MHO~1652, and HH~1002~C are also visible in this image.
    \textbf{(b)}
    The bipolar jet MHO~1639 emerges from a point source, J103653.8-583748, with several knots tracing the asymmetrical flow. 
    Knots extend $\sim$7\arcsec\ to the left of the source, approaching the edge of the cloud. 
    The opposite outflow limb extends nearly the full length of this portion of the cloud, with knots seen $\sim$25\arcsec, or $>0.25$~pc, from the driving source. 
    A black dashed box indicates the zoom area shown in panels (c) -- (e).
    \textit{Bottom:} Images of the structured nebulosity around J103653.8-583748 shown in 
    \textbf{(c)} F200W--F335M, 
    \textbf{(d)} F200W, and 
    \textbf{(e)} F187N. 
    }
    \label{fig:MHOc9}
\end{figure*}
MHO~1639 emerges from a bright point source that lies $\sim$5.5\arcsec\ inside the cloud edge. 
Knots in the southern outflow limb (the left side of the flow in Figure~\ref{fig:MHOc9}) trace a collimated outflow that extends close to the ionized edge of the cloud. 
The northern limb appears wider-angle and more diffuse, as though tracing shocks along the walls of an outflow cavity. 
Additional knots further to the north (to the right side of Figure~\ref{fig:MHOc9}) lie along the same axis, tracing a straight line through the driving source to the southern most knot of the counterflow. 
Bow shocks at the terminus of MHO~1639 are also visible in H$\alpha$ and \pa\ images (the axis connecting these features is shown as a dashed white line in Figure~\ref{fig:MHOc9}a). 
The proper motion of these features suggest that these knots all belong to a single, coherent flow and are discussed in Section~\ref{ss:jets}.  

The immediate environment of the probable MHO~1639 driving source, J103653.8-583748, is intriguing. 
In continuum and \pa\ images, a pillar-like dark cloud obscures the YSO which creates  a halo of emission behind it (see Figure~\ref{fig:MHOc9}~c,d,e). 
Diffuse emission extends from the bright nebulosity with a morphology that suggests it traces the edges of a bubble or cavity near the YSO. 
A bright \mh\ feature seen near J103653.8-583748 that extends perpendicular to the jet may trace the outer wall of such a cavity. 
Overall, the feature bears a strong resemblance to the NGC~1999 reflection nebula as it appears in unpublished \emph{HST} images \citep{noll1999}. 
In both objects, a pillar-like dark cloud is seen amid bright nebulosity. 
However, the dark region in NGC~1999 is a cavity in the cloud \citep{stanke2010}, possibly excavated by the multiple jets that emerge from the nearby V380~Ori system \citep[including the large-scale HH~222 jet, see][]{reipurth2013}. 
In contrast, it appears that a pillar of material obscures the MHO~1639 driving source, J103653.8-583748. 
The possible cavity in this case is offset to the south of the pillar (left in Figure~\ref{fig:MHOc9}) and may have been carved by MHO~1639. 
High-resolution sub-millimeter observations are required to test this hypothesis. 
\\

\textit{MHO~1640:}
%
\begin{figure}
	\includegraphics[width=\columnwidth]{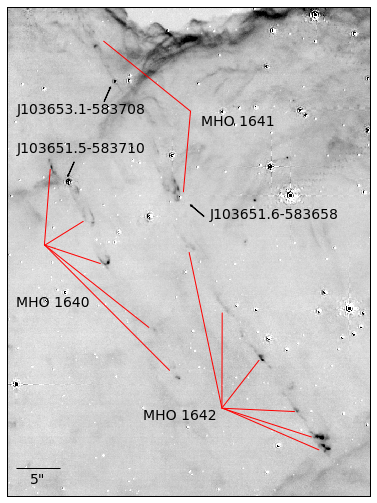}
    \caption{The side-by-side and nearly parallel outflows MHO~1640 and MHO~1641. 
    MHO~1642 is located to the west of MHO~1641 (lower right of the image) although it is unclear if the two are related. 
    }
    \label{fig:MHOc11}
\end{figure}
Two classic bow shocks define the northwest limb of MHO~1640 (pointing to the lower right in Figure~\ref{fig:MHOc11}). 
Two additional shocks beyond these well-defined bow shocks show that this outflow limb extends at least $25\arcsec$, or $>0.25$~pc. 
In contrast, the counterflow to the southeast (upper left in Figure~\ref{fig:MHOc11}) is significantly shorter.  
Only a single arc of H$_2$ emission extends from the probable driving source, J103651.5-583710, with a length of $\sim$3.5\arcsec\ (0.04~pc), or about 15\% of the length of the northwest limb. 
A faint arc of \pa\ emission extends from J103651.5-583710 to the northwest of the driving source, possibly tracing the walls of the outflow cavity. 
\\

\textit{MHO~1641:}
%
Immediately to the north of and nearly parallel to MHO~1640 is another longer flow, MHO~1641 (seen to the right of MHO~1640 in Figure~\ref{fig:MHOc11}). 
Several \mh\ knots trace a collimated outflow extending nearly parallel to MHO~1640. 
Two point sources are seen on the outflow axis, 
J103653.1-583708 and 
J103651.6-583658.  
Tenuous emission extends in either direction from J103653.1-583708, perhaps tracing the origin of the collimated flow while 
J103651.6-583658 sits amid wider-angle emission with a similar morphology to the reflection nebula seen around MHO~1634. 
However, emission from these arcs are not seen at other wavelengths, suggesting that they may instead be bow shocks associated with the outflow. 
\\

\textit{MHO~1642:}
%
Several additional knots extend to the northwest of MHO~1641 (extending to the lower right in Figure~\ref{fig:MHOc11}). 
These knots trace a remarkably straight line of emission that imply a collimated outflow axis that is slightly offset from MHO~1641. 
Tenuous \mh\ emission may trace one side of an outflow cavity, while a few brighter knots may trace shocks within the jet itself. 
Future proper motion measurements will help clarify the relationship between MHO~1642 and other \mh\ and H-emitting knots in the region (see Section~\ref{ss:jets}). 
\\

\textit{MHO~1643:}
%
\begin{figure*}
	\includegraphics[width=0.4805\textwidth]{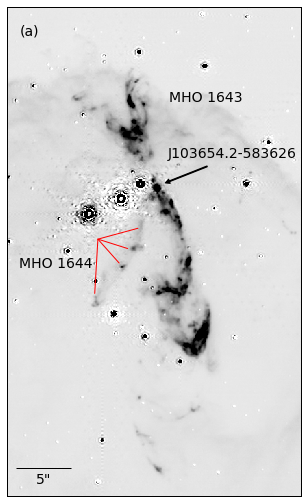}
	\includegraphics[width=0.465\textwidth]{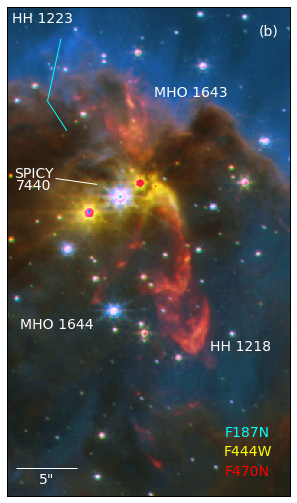}
    \caption{
    \textbf{(a)} The brightest and most spectacular outflow in NGC~3324 is MHO~1643. 
    MHO~1643 extends into the cloud with two prominent bow shocks and several additional knots ahead of and slightly off-axis from the bow shocks. 
    The flow opens toward the cloud edge with \mh\ emission ending at the ionization front. 
    A second outflow, MHO~1644, emerges from nearly the same origin as MHO~1643. 
    MHO~1644 appears monopolar, with a single outflow limb curving away from MHO~1643. 
    \textbf{(b)} Color image with strong \mh\ emission from MHO~1643 and MHO~1644 seen in red. \pa\ emission from HH~1223 is seen in blue as arcs of emission emerging from SPICY~7440 and an additional arc 
    outside the cloud edge. 
    }
    \label{fig:MHOc13}
\end{figure*}
The most dramatic H$_2$ flow seen in the \emph{JWST} images is MHO~1643. Twin bow shocks stand out in color images, and trace back to a small cluster of stars that lies close to the cloud edge. 
Continuum-subtracted H$_2$ images (see Figure~\ref{fig:MHOc13}) reveal shocks and diffuse emission tracing one wall of the outflow cavity from the probable driving source, J103654.2-583626, to the first bow shock. 
Beyond the first bow shock, diffuse emission hints at the outflow cavity created by the bow shock further northwest (down in Figure~\ref{fig:MHOc13}). 
Fainter shock-like structures beyond the two prominent bow shocks hint that there have been multiple outbursts from this source. 
On the opposite side of the driving source, 
MHO~1643 opens in a V-shape giving the flow an hour-glass like shape around the YSO. 
The outflow cavity points toward the cloud edge with molecular emission ending at the ionization front.

MHO~1643 is offset to the north of three bright stars. 
A fourth fainter star can be identified at the apparent origin of the flow in the \emph{JWST} images, J103654.2-583626.  
Together, this collection of stars appears to blow a bubble to the southeast of the MHO~1643 flow (see Figure~\ref{fig:MHOc13}b).
Bright emission in both narrowband (F187N) and broadband (F200W) images traces arcs of emission reminiscent of billowing smoke. 
\citet{smith2010} identified a few shock-like features in \ha\ from this region as candidate jet HH~c-1 (now HH~1223; see Figure~\ref{fig:MHOc13} and Section~\ref{sss:pmf3}). 
The motion of features detected in both \pa\ and H$\alpha$ indicate that the feature is in fact part of an outflow, but that it is separate from MHO~1643 (see Section~\ref{ss:jets}). 
Interior to these arcs, there is little emission suggesting that the nascent cluster is creating a cavity. 
The bright YSO at the base of this feature is SPICY~7440, a flat spectrum YSO that is $\sim$3.5\arcsec\ to the southwest (left in Figure~\ref{fig:MHOc13}) of the origin of MHO~1643. 
\\

\textit{MHO~1644:}
%
A second, monopolar flow emerges from the same cluster of stars as MHO~1643 (see Figure~\ref{fig:MHOc13}). 
MHO~1644 consists of a chain of three bright \mh\ knots that are connected by an arc of diffuse H$_2$ emission. This outflow seems to originate at almost the same place as MHO~1643, then bends to the southwest (lower left in Figure~\ref{fig:MHOc13}). 
It is unclear which source drives MHO~1644 and how close it is to the MHO~1643 driving source. 
In the future, proper motion measurements will clarify the relationship between the outflows and YSOs in this region. 
\\

\textit{MHO~1645:}
%
\begin{figure}
	\includegraphics[width=\columnwidth]{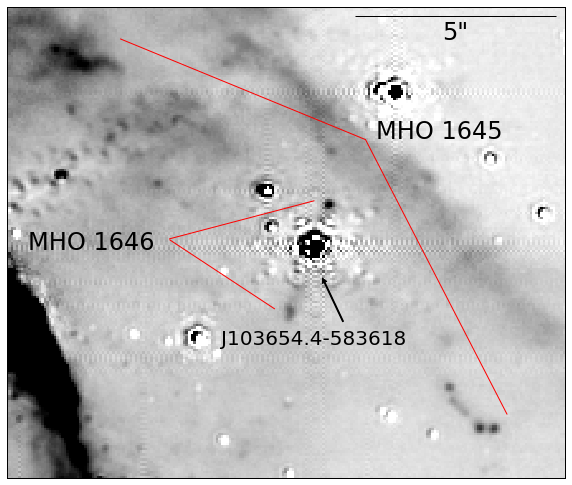}
    \caption{MHO~1645 and its probable driving source, J103654.4-583618, lie $\sim$3\arcsec\ inside the edge of the cloud.
    The bipolar outflow propagates nearly parallel to the cloud edge. 
    A second outflow, MHO~1646, appears to emerge from the same source with an orientation nearly perpendicular to MHO~1645. 
    }
    \label{fig:MHOc15}
\end{figure}
MHO~1645 is a bipolar outflow that emerges from a prominent star, J103654.4-583618, that lies just inside the ionization front (see Figure~\ref{fig:MHOc15}). 
Some diffuse emission to the southeast (upper left in Figure~\ref{fig:MHOc15}) of the driving source seen in the F090W images 
may trace reflected light at the wide-angle base of the flow. 
\\

\textit{MHO~1646:}
%
Two additional \mh\ knots are seen on either side of J103654.4-583618 and may trace a second outflow, MHO~1646, that is oriented nearly perpendicular to MHO~1645. 
The eastern knot (pointing to the lower left in Figure~\ref{fig:MHOc15}) extends outside the PSF subtraction residuals, with faint emission extending back toward the point source. 
Similar to MHO~1637, a bright knot on the opposite side of the star may trace the counterflow, although this is less certain as the knot may be confused with subtraction residuals.  
\\

\textit{MHO~1647:}
%
\begin{figure}
	\includegraphics[width=\columnwidth]{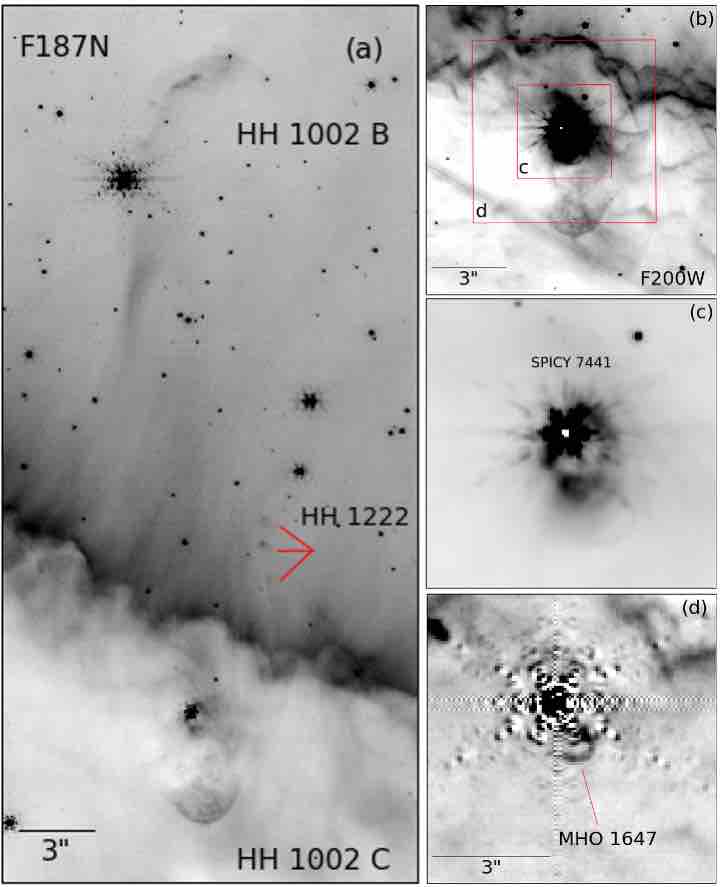}
	\caption{
    \textbf{(a)} \pa\ image of HH~1002~B and HH~1002~C, an HH jet identified by \citet{smith2010} that is associated with MHO~1647. The neighboring jet, HH~1222, is also labeled. 
    \textbf{(b)} An F200W image showing the HH~1002~C bow shock and boxes indicating the zoomed in regions shown in panels (c) and (d). 
    \textbf{(c)} More structure in the  MHO~1647 bow shock is apparent in the F200W image as there is less contamination from the diffraction  spikes from the YSO. 
    \textbf{(d)} \mh\ emission traces a well-defined bow shock that emerges from the YSO SPICY~7441. 
     }
    \label{fig:MHOc16}
\end{figure}
Continuum-subtracted H$_2$ images reveal a single bow shock emerging to the northwest of the SPICY~7441 YSO (pointing to the lower right in Figure~\ref{fig:MHOc16}). 
Any additional structure interior to this shock is confused with strong subtraction residuals around the bright YSO. 
More structure can be identified in the F200W image where emission likely traces a combination of \mh\ 2.12~\micron\ and Br$\gamma$ 2.16~\micron\ emission. 
Some faint \pa\ emission is also seen from this feature. 

A second bow shock located $\sim$3\arcsec\ further from the YSO lies along the same axis. 
The feature traces a larger arc with prominent emission in the \pa\ and F200W images. 
The same morphology was seen in H$\alpha$ images and identified as HH~1002~C by \citet{smith2010}. 
We use the H$\alpha$ and \pa\ images to measure the proper motion of HH~1002~C; see discussion in Section~\ref{ss:jets}. 
\\

\textit{MHO~1648:}
%
\begin{figure}
	\includegraphics[width=\columnwidth]{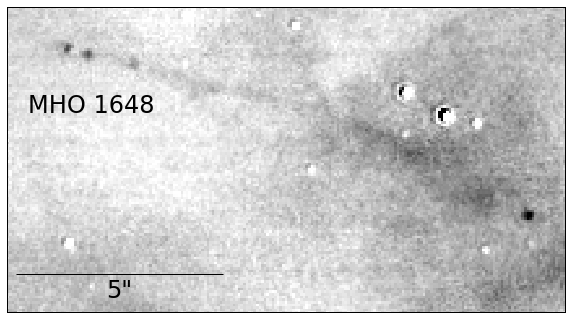}
    \caption{Faint, curved emission traces MHO~1648. No obvious driving source lies within or near the flow.  }
    \label{fig:MHOc17}
\end{figure}
MHO~1648 (see Figure~\ref{fig:MHOc17}) is a tenuous arc of emission. 
Four knots interspersed in the flow likely trace shocks. 
There is no candidate YSOs near the outflow axis, so it is unclear what source might drive the flow.   
\\

\textit{MHO~1649:}
%
\begin{figure}
	\includegraphics[width=\columnwidth]{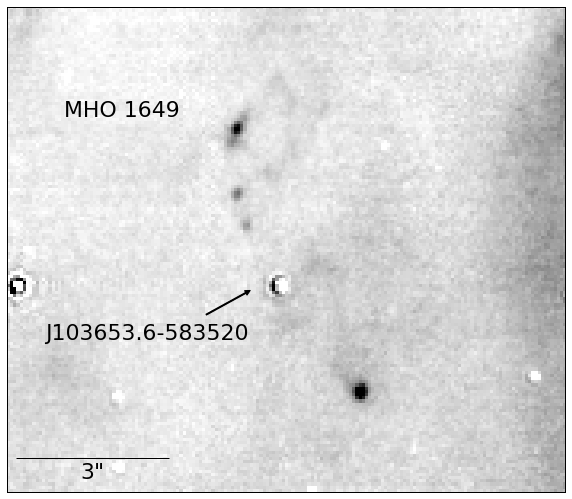}
    \caption{The small, arced outflow MHO~1649 emerges from a star, J103653.6-583520, at origin of the flow. }
    \label{fig:MHOc18}
\end{figure}
A series of four shocks with diffuse \mh\ emission between them trace MHO~1649 (see Figure~\ref{fig:MHOc18}). 
Three knots curve to the southeast (toward  the top of Figure~\ref{fig:MHOc18}) of the probable driving source, J103653.6-583520.  
Only a single knot has been identified in the counterflow. 
The outflow axis appears to  bends slightly toward the northeast (upper right in Figure~\ref{fig:MHOc18}). 
\\

\textit{MHO~1650:}
%
\begin{figure}
	\includegraphics[width=\columnwidth]{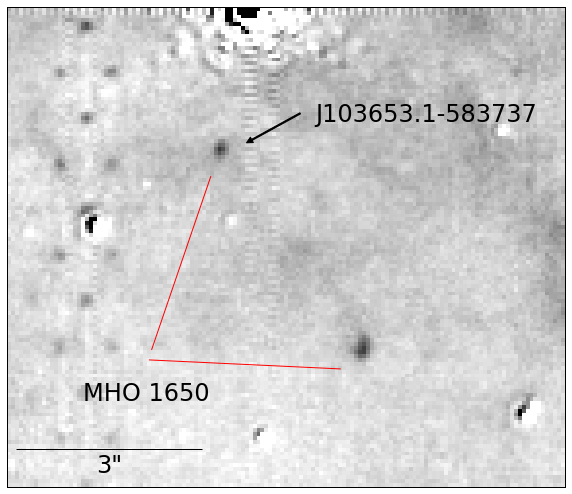}
    \caption{Two knots of \mh\ emission make up MHO~1650. The first knot is coincident with a point source, J103653.1-583737. The bow-shock shape of the second knot suggests that the outflow propagates into the cloud. }
    \label{fig:MHOc19}
\end{figure}
Two knots make up MHO~1650 (see Figure~\ref{fig:MHOc19}). 
One of the two knots, seen in the lower right of Figure~\ref{fig:MHOc19}, has a rounded shape, reminiscent of a small bow shock.
Tenuous emission extending from the shock hints at emission from the walls of an outflow cavity. 
The other bright spot of H$_2$ emission is coincident with a point source, J103653.1-583737, that likely drives the monopolar flow. 
\\

\textit{MHO~1651:}
%
\begin{figure}
	\includegraphics[width=\columnwidth]{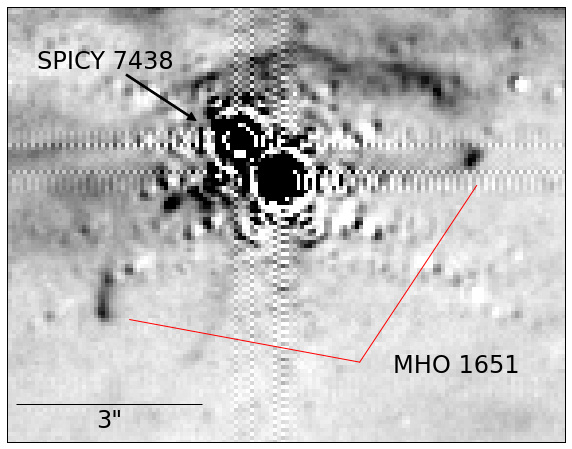}
    \caption{Two knots of \mh\ emission on either side of the YSO SPICY~7438 are MHO~1651. 
    Like MHO~1645, this outflow lies $\sim$4\arcsec\
    within the cloud edge and extends nearly parallel to the ionization front. }
    \label{fig:MHOc20}
\end{figure}
Two knots on either side of the YSO SPICY~7438 trace MHO~1651 (see Figure~\ref{fig:MHOc20}). 
The system lies just inside and nearly parallel to the edge of the irradiated cloud. 
Tenuous emission from the two shocks extends back toward the YSO, tracing a curved flow that bends away from the ionized cloud edge. 
\\

\textit{MHO~1652:}
%
\begin{figure}
	\includegraphics[width=\columnwidth]{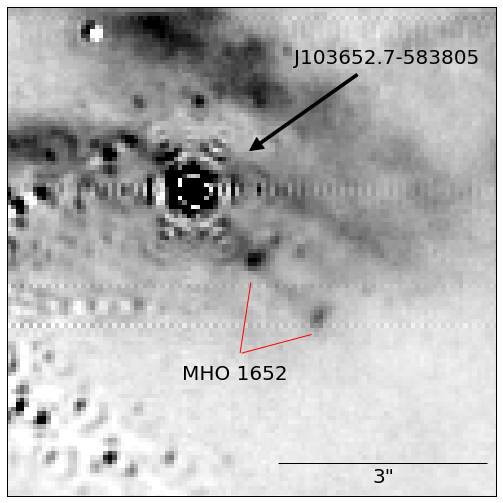}
    \caption{Two knots of \mh\ emission trace the monopolar MHO~1652 as it emerges from J103652.7-583805, a bright star that resides just inside the ionization front. The outflow propagates into the cloud with no evidence for a counterflow seen in \pa.  }
    \label{fig:MHOc21}
\end{figure}
MHO~1652 consists of two \mh\ knots propagating from a point source, J103652.7-583805 (see Figure~\ref{fig:MHOc21}). 
The driving source lies just inside a ridge of the ionization front, so the counterflow may propagate outside the cloud where it is rapidly dissociated and ionized in the H~{\sc ii} region. 
No counterflow is evident in continuum-subtracted \mh\ images or in \pa\ emission outside the cloud at this location. 
\\

\textit{MHO~1653 and MHO~1654:}
%
\begin{figure}
	\includegraphics[width=\columnwidth]{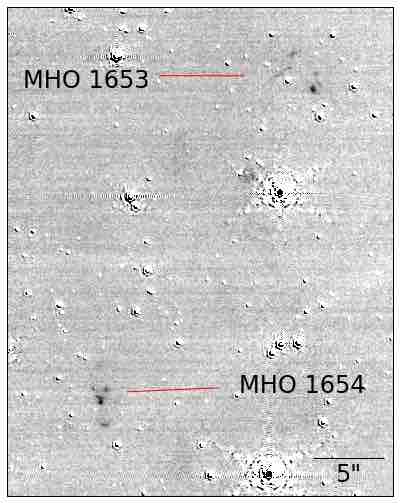}
    \caption{Two regions of small \mh\ knots are MHO~1653 and MHO~1654. Like MHO~1631, these knots are seen in the H~{\sc ii} region, far from the ionized cloud edge. No jet body or candidate driving source connect the two \mh\ features.   }
    \label{fig:MHOc22}
\end{figure}
Two collections of small \mh\ knots are MHO~1653 and MHO~1654 (see Figure~\ref{fig:MHOc22}).  
These knots lie just below MHO~1631 and also likely reside behind the NGC~3324 H~{\sc ii} region.  
No candidate YSOs are detected near either knot and no jet body demonstrates that the two features are physically related. 
Both features trace a shock-like morphology. 
However, the apex and wings of the purported shocks suggest that these features may not have a common origin. 
The rounded leading edge of MHO~1654 (closest to the bottom in Figure~\ref{fig:MHOc22}) suggests that it is a bow shock followed by a chain of knots. 
The axis implied by these features does not intersect MHO~1653. 
The knots and arcs of MHO~1653 provide fewer hints at their possible direction of motion, so their origin is unclear.

%
\begin{table*}
	\centering
	\caption{H$_2$ jets in NGC~3324. }
	\label{t:jets}
	\begin{tabular}{lccrccl} 
		\hline
		Name & R.A. & Dec. & P.A. & YSO & Stage & Comment \\
		\hline
		MHO~1631 & 10:37:05.8 & $-$58:37:30 & 52$^{\circ}$ & ... & ... & in the H~{\sc ii} region   \\ 
		MHO~1632 & 10:36:42.7 & $-$58:38:05 & 108$^{\circ}$ & J103642.3-583804 & ... & \\
		MHO~1633 & 10:36:48.2 & $-$58:38:17 & 179$^{\circ}$ & J103648.0-583819 & ... & \\ 
		MHO~1634 & 10:36:47.2 & $-$58:38:10 & 51$^{\circ}$ & J103647.3-583810 & Class~I & SPICY~7423  \\
		MHO~1635 & 10:36:46.7 & $-$58:38:05 & 57$^{\circ}$ & J103646.7-583805  & ... & \\
		MHO~1636 & 10:36:51.0 & $-$58:37:52 & 119$^{\circ}$ & J103651.5-583754 & ... & \\
		MHO~1637 & 10:36:50.0 & $-$58:37:51 & 111$^{\circ}$ & J103650.5-583752 & ... & \\
		MHO~1638 & 10:36:51.4 & $-$58:37:50 & 154$^{\circ}$ & J103651.4-583748  & ... & \\ 
		MHO~1639 & 10:36:53.9 & $-$58:37:39 & 6$^{\circ}$ & J103653.8-583748 & ... & HH~1003~A and HH~1221 trace associated bow shocks \\
		MHO~1640 & 10:36:51.5 & $-$58:37:10 & 132$^{\circ}$ & J103651.5-583710 & ... & \\
		MHO~1641 & 10:36:53.3 & $-$58:37:10 & 131$^{\circ}$ & ... & ... & two possible driving sources: J103653.1-583708 and J103651.6-583658 \\
		MHO~1642 & 10:36:49.6 & $-$58:36:44 & 135$^{\circ}$ & ... & ... & associated with HH~1224? \\ 
		MHO~1643 & 10:36:54.2 & $-$58:36:27 & 109$^{\circ}$ & J103654.2-583626 & ... & associated with HH~1218\\
		MHO~1644 & 10:36:53.3 & $-$58:36:28 & 73$^{\circ}$ & ... & ... & looks monopolar \\ 
		MHO~1645 & 10:36:54.1 & $-$58:36:14 & 151$^{\circ}$ & J103654.4-583618 & ... & \\ 
		MHO~1646 & 10:36:54.2 & $-$58:36:19 & 83$^{\circ}$ & J103654.4-583618 & ... & \\ 
		MHO~1647 & 10:36:53.9 & $-$58:37:19 & ... & J103654.0-583720 & Class~I & SPICY~7441; bow shock near YSO; part of HH~1002~C \\
		MHO~1648 & 10:36:58.0 & $-$58:34:48 & 173$^{\circ}$ & ... & ... & no obvious driving source  \\
		MHO~1649 & 10:36:53.6 & $-$58:35:20 & 128$^{\circ}$ & J103653.6-583520 & ... & \\
		MHO~1650 & 10:36:52.9 & $-$58:37:36 & 139$^{\circ}$ & J103653.1-583737 & ... & \\
		MHO~1651 & 10:36:53.3 & $-$58:37:54 & 35$^{\circ}$ & J103653.3-583754 & uncertain & SPICY~7438\\
		MHO~1652 & 10:36:52.7 & $-$58:38:05 & 148$^{\circ}$ & J103652.7-583805 & ... &  \\
		MHO~1653 & 10:37:04.8 & $-$58:37:15 & ... & ... & ... & in the H~{\sc ii} region \\
		MHO~1654 & 10:37:01.5 & $-$58:37:24 & ... & ... & ... & in the H~{\sc ii} region, near MHO~1653  \\
		\hline
	\end{tabular}
\end{table*}

\subsection{Kinematics of atomic jets}\label{ss:jets}

All of the irradiated HH jets and candidate jets identified in the 2006 H$\alpha$ images from \emph{HST} by \citet{smith2010} are also visible in the new 2022 \pa\ images from \emph{JWST}. 
We can therefore measure their proper motions to determine the speed and direction of the outflowing material, 
distinguish separate outflows, constrain shock velocities 
and feedback energies, and identify potential driving sources. These kinematic measurements also confirm that, except for the wispy filaments along the sides of HH~1002~B and HH~1003~B, each of the previously identified jet candidates are 
parts of outflows, and 
not merely static filamentary structures in the H~{\sc ii} region. 

Using the IRAF\footnote{IRAF was distributed by the National Optical Astronomy Observatory (now NOIRLab), operated by the Association of Universities for Research in Astronomy, Inc., under cooperative agreement with the National Science Foundation.} tasks GEOMAP and GREGISTER, 338 stars in common served as tie points to align the \emph{HST}/ACS F658N H$\alpha$+[N~II] mosaic image to the \emph{JWST}/NIRCam short wavelength channel F187N \pa\ image with a final calculated image scale of $30.92''$ pixel$^{-1}$. Quadratic fits in X and Y were used to map the ACS data to the NIRCam image scale and orientation, with rms residuals of $\sim$0.7 pixels (or $\sim$22 mas). About 5\% of the stars were rejected during the fitting process due to high proper motions during the 16-yr time baseline.\footnote{A similar exercise aligning two ACS F475W images obtained about a decade apart of the young supernova remnant 1E~0102.2-7219 in the Small Magellanic Cloud yielded rms residuals of $\sim$0.08 ACS pixels or 4 mas using 328 stars and three background galaxies as tie points. With the stellar tie points being mostly at the distance of the SMC, the excellent image alignment is essentially limited by variability in the instrument optical distortions and the \emph{HST} pointing jitter. Thus we believe the larger residuals in aligning the 2006 ACS F658N image mosaic to the 2022 NIRCam F187N mosaic derive from intrinsically larger scatter in the stellar tie point positions due to measurable proper motions at the distance of NGC~3324, and possibly to slight errors in stitching the NIRCam image mosaic together. } \citep[See][for a related discussion of multi-epoch image alignment in the Carina Nebula region.]{reiter2015_hh900}

At a distance to NGC~3324 of 2.3~kpc and a time baseline of $5.1 \times 10^8$~s, one pixel of motion corresponds to 20.8~\kms. Typical transverse motions for outflow features range from $<2$ to $\sim$7~pixels or $\sim$25 -- 150~\kms, consistent with protostellar jet velocities observed elsewhere in the Carina Nebula region \citep[cf.~Figure~28 in][]{reiter2017}. All proper-motion measurements were estimated `by eye' by comparing the H$\alpha$ and \pa\ images on the computer display. However, many of the well-delineated features were measured with the proper motion code used previously in \citet{mor01}, \citet{hartigan2005}, and \citet{kiminki2016}, 
based on an algorithm described in \citet{cur96}.
The estimated measurement uncertainty for individual features is on the order of $\pm 1$~pixel or $\pm \sim$20~\kms.

%
%
\begin{figure*}
\includegraphics[width=\textwidth]{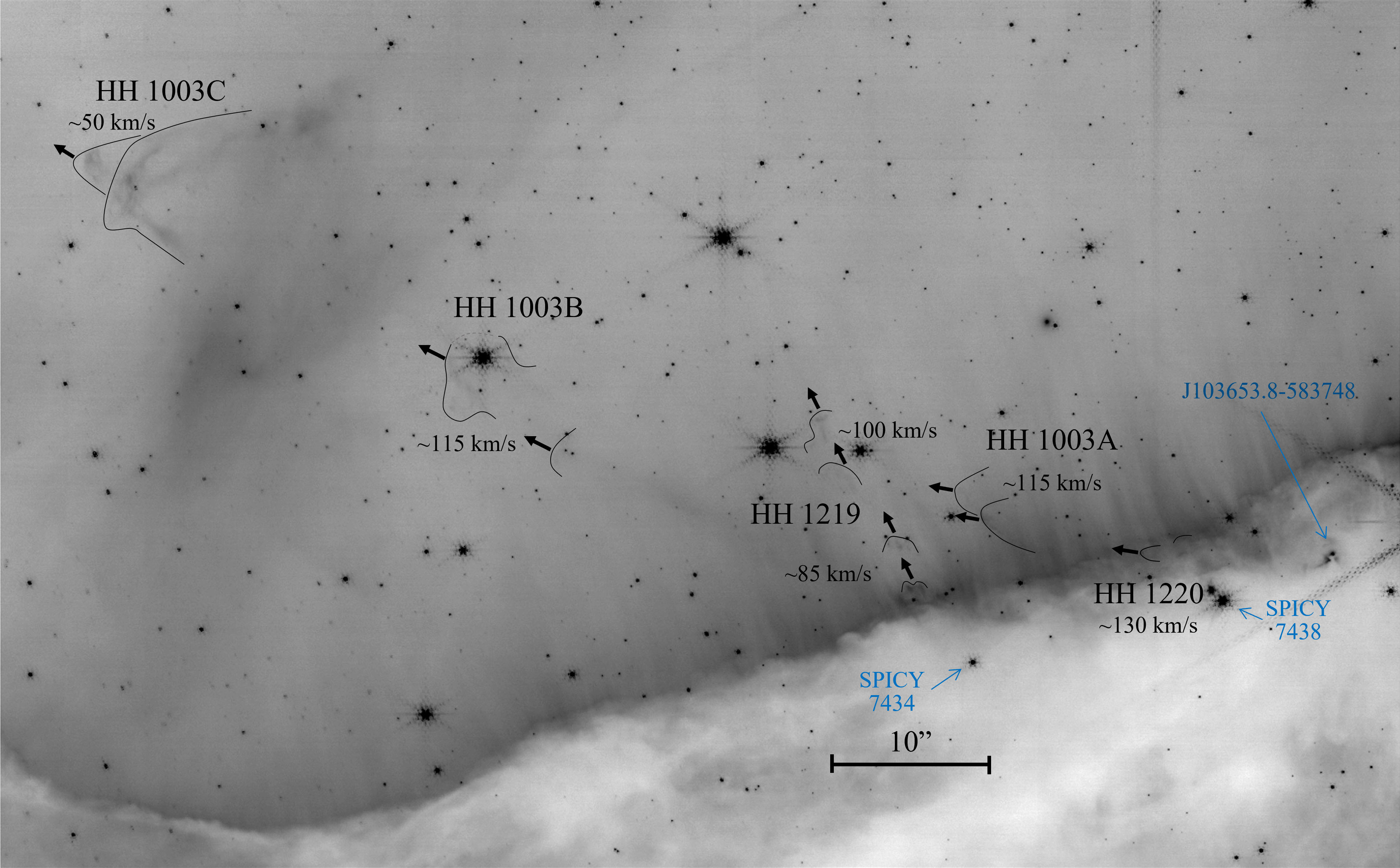}
	\caption{ 
    Detail of PM Field 1 from the region outlined in the upper panel of Figure~\ref{fig:overview}. H-emitting outflow features with measured proper motions are shown on the 2022 JWST NIRCam F187N image. All labeled features are also visible in the HST ACS F658N image from 2006. In this field, shock features in the HH~1003A, B \& C complexes described by \citet{smith2010} are delineated with thin lines and labeled. Transverse speeds and directional arrows accompany the labels. The kinematic measurements reveal that multiple optical outflows are superposed. Several potential YSO driving sources are noted in blue text and arrows.}
    \label{fig:pms_field1}
\end{figure*}
%
%
\begin{figure}
\includegraphics[width=\columnwidth]{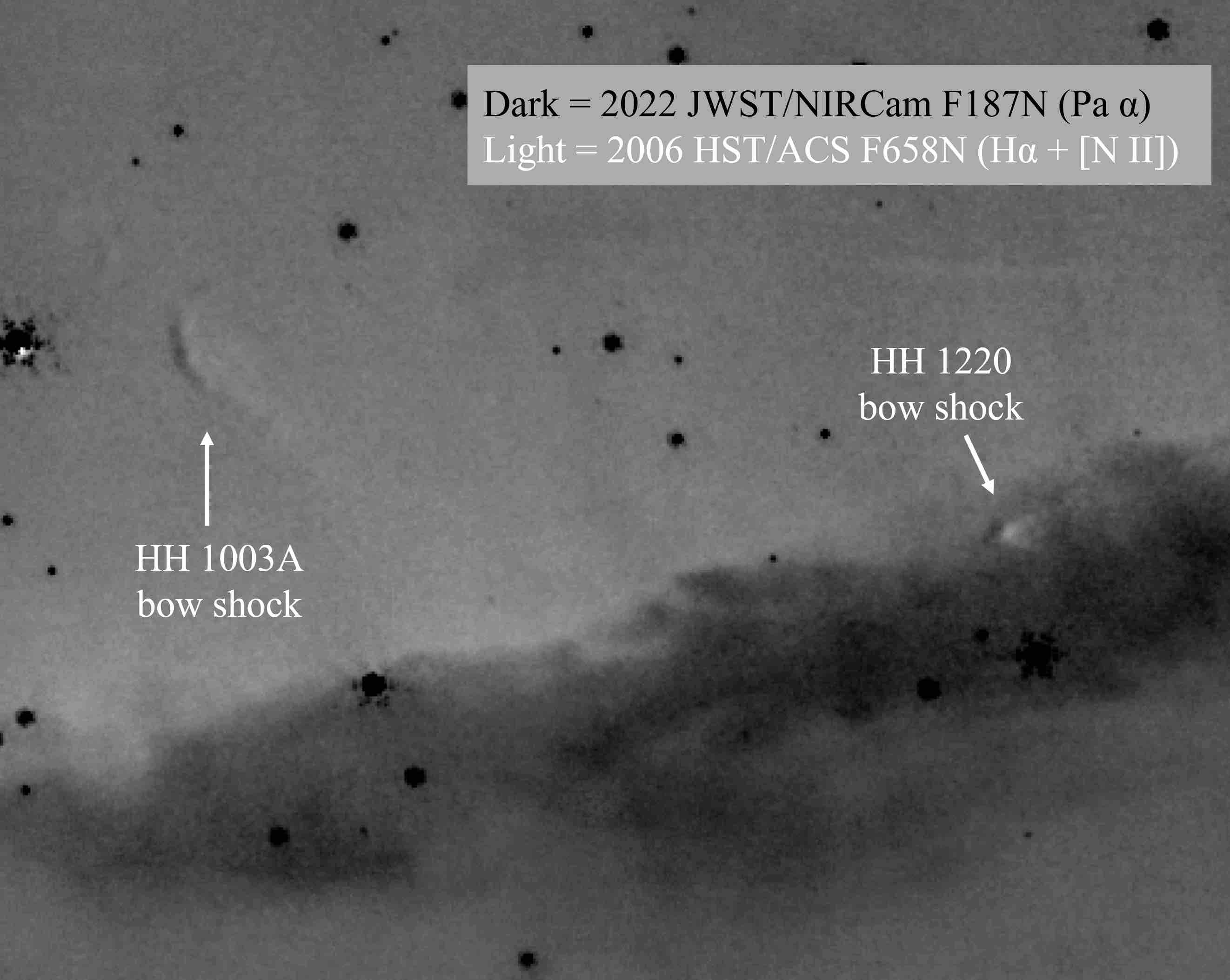}
        \caption{
    H$\alpha$ minus \pa\ difference image of the region around the prominent bow shocks HH~1003~A and HH~1220. Dark leading edges indicate the motion of the shock from 2006 to 2022, for these two objects $\sim$5-7 pixels.
        }
        \label{fig:pms_diff}
\end{figure}
%
%
\begin{figure*}
\includegraphics[width=\textwidth]{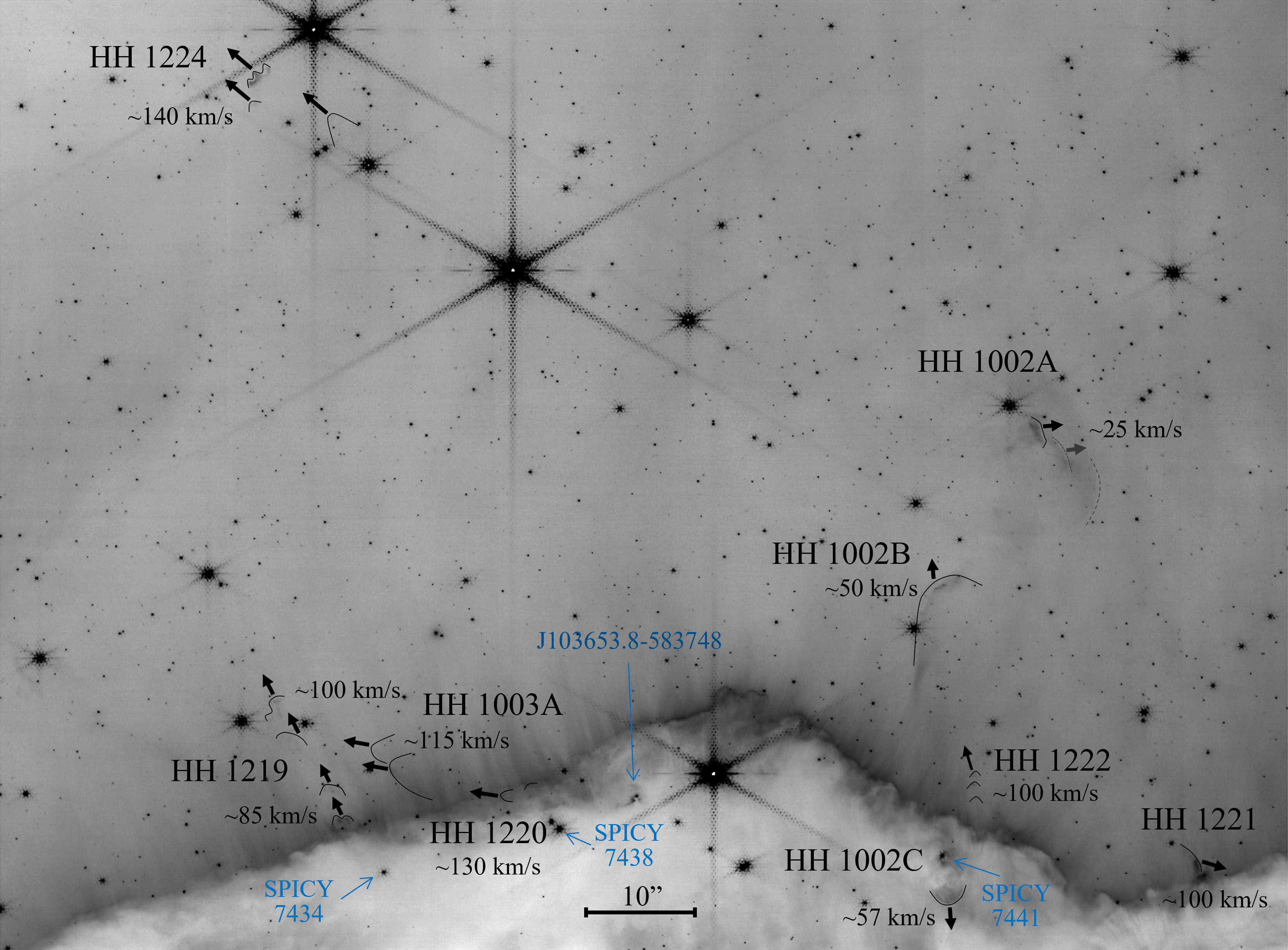}
    \caption{ 
    Detail of PM Field 2 from the region outlined in the upper panel of Figure~\ref{fig:overview}. H-emitting outflow features with measured proper motions are shown on the 2022 JWST NIRCam F187N image. All labeled features are also visible in the HST ACS F658N image from 2006. In this field, shock features in the HH~1003A \& HH~1219 complexes are repeated from PM Field~1 and shown in context with the HH~1002A, B, \& C, and HH~1224 structures described by \citet{smith2010}. Transverse speeds and directional arrows accompany the feature labels. The kinematic measurements reveal that multiple optical outflows are superposed. Several potential YSO driving sources are noted in blue text and arrows.}
    \label{fig:pms_field2}
\end{figure*}

\begin{figure}
\includegraphics[width=3.37in]{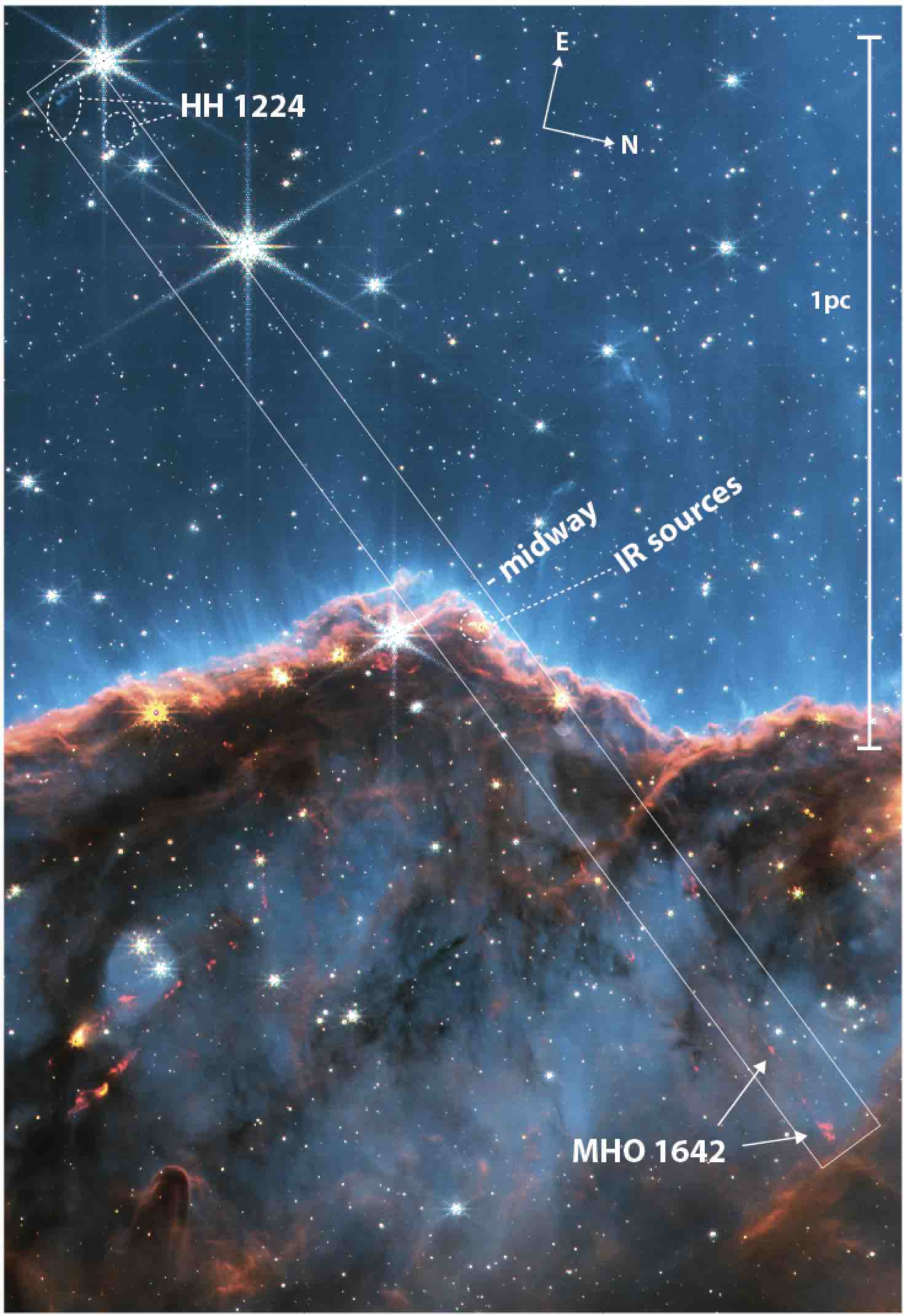}  
    \caption{Color image (same color coding and orientation as Fig.~\ref{fig:overview}) indicating a possible parsec-scale outflow that includes both optical HH objects and \mh-emitting knots.  The large tilted white box follows the putative axis of the large bipolar outflow, with HH~1224 at the upper left end and MHO~1642 at the lower right end.  Near the halfway point between these two (the middle of the box is marked ``midway'') we note three IR sources, one of which could be a protostar that potentially drives the large outflow. If this is a single bipolar outflow, the full projected length is $\sim$1.8 pc.}
    \label{fig:bigjet}
\end{figure}

%
%
\begin{figure*}
    \includegraphics[width=\textwidth]{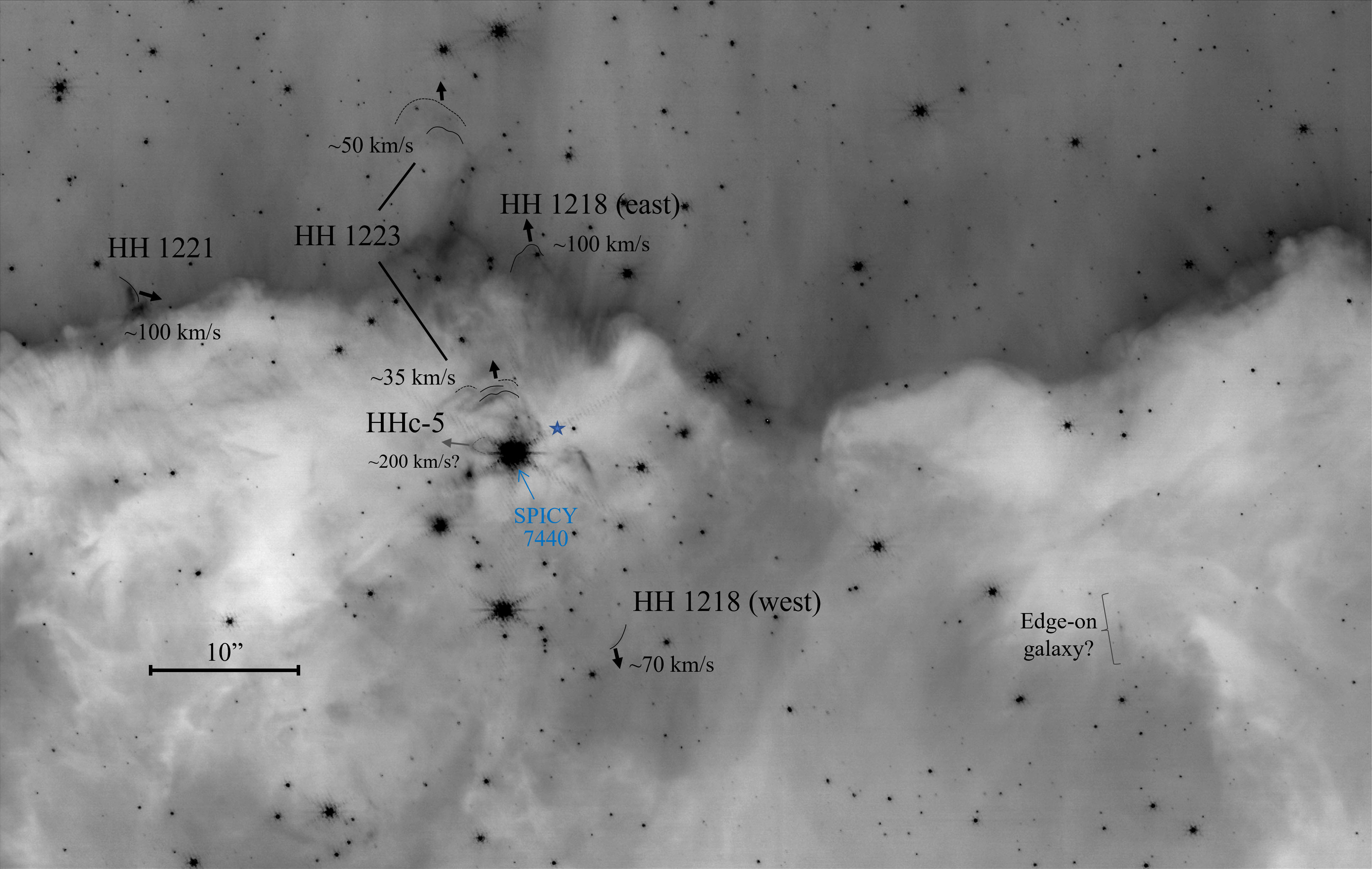}
	\caption{ 
    Detail of PM Field 3 from the region outlined in the upper panel of Figure~\ref{fig:overview}. H-emitting outflow features with measured proper motions are shown on the 2022 JWST NIRCam F187N image. All labeled features are also visible in the HST ACS F658N image from 2006, except for the candidate edge-on galaxy in the lower-right. In this field, shock features in the HH~1221 feature is repeated from PM Field~2 and shown in context with the HH~1223 structures described by \citet{smith2010}. Transverse speeds and directional arrows accompany the feature labels. As in PM Fields 1 and 2, the kinematic measurements reveal that multiple optical outflows are superposed. The SPICY~7440 YSO driving source of HH~1223 is noted in blue text. A blue star symbol just to the right and above SPICY~7440 marks the approximate position of the driving source of HH~1218/MHO~1643.}
    \label{fig:pms_field3}
\end{figure*}
The kinematics of individual outflows and jets are described below in the context of larger fields of regard (labeled as PM Fields 1-3 in Figure~\ref{fig:overview}a). 
We adopt the nomenclature of 
\citet{smith2010} who provided an overview of HH objects and jet-driven features in the portion of NGC~3324 captured with the \emph{HST} (see Figure~\ref{fig:overviewBIG}c).  
However, we note instances where features within a single HH designation may represent more than one outflow, or where features with different HH designations appear kinematically connected.
In addition, several of the optical HH jets are associated with \mh\ counterparts.
The kinematic measurements of H-emitting features are depicted in Figures~\ref{fig:pms_field1}, \ref{fig:pms_field2}, and \ref{fig:pms_field3}, corresponding to the overlapping PM Fields 1--3 shown as dashed boxes in Figure~\ref{fig:overview}a.

\subsubsection{PM Field 1}
\label{sss:pmf1}
\textit{HH~1003:}
The field shown in Figure~\ref{fig:pms_field1} includes several optical shock structures that \citet{smith2010} labeled HH~1003~A, B, \& C. The proper motions 
-- denoted in Figure~\ref{fig:pms_field1} by the transverse speeds in \kms\ and arrows -- reveal that the HH~1003 complex may comprise three separate outflows. 
The star (bright in the F187N filter) projected in the middle of the forward HH~1003~B shock features is not identified as a YSO in the SPICY catalog and appears to be a chance superposition. 
The HH~1003~B shocks travel at $\sim$115~\kms\ in the same direction as the outer, slow-moving HH~1003~C complex and thus are part of the same outflow.

The HH~1003~C bow shock overall is slow-moving, with a measured proper-motion speed of $\sim$50~\kms. 
However, a few small filaments within the bow shock near the apex appear to move faster ($\sim$80~\kms), though this could reflect changing caustics along our line of sight through the hydrodynamically evolving structure rather than real radiative shock wave motions. 
At $\sim$50~\kms, the time for the HH~1003~C bow shock to traverse 
$\gtrsim$1\arcmin\ 
(projected distance of $\sim$0.7~pc) 
from the cloud edge near HH~1003~A to its current position is $\sim$14,000~years. 
Such a timescale is consistent with those estimated for other outer bow shocks in parsec-scale HH outflows \citep[e.g.,][]{devine1997,stanke1999}. 
However, given the speed of HH~1003~B ($\sim$115~\kms) in its wake, we could infer that HH~1003~C probably had a higher velocity in the past and has slowed during its transit through the H~{\sc ii} region. If it had an average transverse speed of $\sim$100~\kms, the transit time is $\sim$7000 years. At a projected distance spanning 30\arcsec\ -- 40\arcsec\ ($\sim$0.33--0.45~pc) from the cloud edge and traveling $\sim$115~\kms, the HH~1003~B group transit time is $\sim$2800--3800 years, and the time between the B \& C major outbursts is at least several thousand years.
Looking back along the direction of propagation towards the molecular cloud, there are a number of IR sources that could be driving these outflow features. 
SPICY~7438, the star just to the right (north) of HH~1220, lies roughly along the flow axis connecting HH~1003~B \& C. 
However, no other optical outflow features emanate from this source. 
MHO~1651 emerges from SPICY~7438 with an axis nearly perpendicular direction to the HH~1003~B \& C flow axis (see Figure~\ref{fig:MHOc20} and Section~\ref{ss:H2jets}) and is clearly a separate flow.

The HH~1003~A features present a more complicated velocity field, and we sub-divide these into two groups. 
HH~1003~A was described in \citep{smith2010} as the probable driving `jet' of the B \& C features, but this appears instead to be a double-bow shock moving in a direction somewhat closer to the cloud edge. In Figure~\ref{fig:MHOc9} we noted the apparent association of HH~1003~A with MHO~1639 and a candidate driving source; we return to this in the discussion of HH~1221 below. 
Figure~\ref{fig:pms_diff} displays the H$\alpha$ minus \pa\ difference image that shows how the HH~1003~A shocks have moved between 2006 and 2022. 
Also apparent in the right-hand side of the difference image is a small, well-defined bow shock, HH~1220. This feature has an even higher transverse velocity than HH~1003~A but is traveling along a line closer to the cloud edge.

Nearby is the second sub-group, a collection of shock features that define an outflow emerging in a direction nearly perpendicular to the cloud edge into the H~{\sc ii} region, quite distinct from the general direction of the HH~1003~A, B, \& C features. 
These knots clearly trace a separate jet that is now identified as HH~1219.  
An embedded source, SPICY~7434, lies just inside the cloud edge along the outflow axis and is the likely driving source.
The kinematic ages of the four HH~1219 features noted in Figure~\ref{fig:pms_field1} are 770, 1100, 1650, and 2000 years from closest to furthest, respectively.
\\

\textit{HH~1220:}
HH~1220 is a compact bow shock that lies along 
the bright ionization front at the cloud edge (see Figures~\ref{fig:pms_field1} and \ref{fig:pms_field2}). 
The bow-shaped arc seen in \pa\ and H$\alpha$ is confused with the undulations of the ionization front in single-epoch images, however the feature clearly moves between epochs, as shown in the subtraction image in Figure~\ref{fig:pms_diff}. The feature has a relatively high transverse velocity ($\sim$130~\kms), consistent with it being a bow shock driven by an unseen jet. The HH~1220 direction of propagation runs slightly closer to the cloud edge than HH~1003~A, and it does not lie along the line in Figure~\ref{fig:MHOc9} from the HH~1003~A bow shock apex through the MHO~1639 outflow and driving source to the HH~1221 bow shock. Therefore, we hesitate to declare HH~1220 part of the MHO~1639 outflow, however its proximity is certainly suggestive. Future proper-motion and radial-velocity measurements could help to clarify the relationship.

\subsubsection{PM Field 2}
\label{sss:pmf2}
In Figure~\ref{fig:pms_field2}, we present PM Field 2, which includes the HH~1003~A \& HH~1219 H-emitting shock structures in the lower-left, along with HH~1224 \citep[candidate HH~c-2 in][]{smith2010} in the upper-left and HH~1002 features on the right.  
\\

\textit{HH~1224:}
Features in the candidate jet HH~c-2 identified by \citet{smith2010} have high proper motions and move as an ensemble in a direction implied by the arcuate morphology. 
This confirms it as an HH object; it is now designated as HH~1224. 
The outflow driving source is not immediately obvious but would be located towards the lower-right portion of the field shown or deeper into the cloud. 
The kinematics show that neither HH~1219, HH~1002~B, nor HH~1222
propagate towards HH~1224, further implying that these are all distinct outflows.

Taking a larger-scale view in Figure~\ref{fig:bigjet}, there is a tantalizing alignment between the flow axis implied by the motion of the HH~1224 features and the \mh\ knots of MHO~1642. 
A cluster of three IR-bright stars (at R.A.=10:36:54.8 and Dec=-58:37:32) is located at the midpoint between HH~1224 and MHO~1642; further analysis is required to determine if any of these are YSOs. 
If confirmed as a single flow, this would be one of the longest bipolar flows in the region with a total length of $\sim$1.8~pc. 
Moving at a transverse speed of $\sim$140 \kms and assuming one of the IR stars is the driving source, the kinematic age of the HH~1224 features is $\sim$6000 years.
\\

\textit{HH~1002:}
Like HH~1003 in PM Field 1, HH~1002 appears to comprise several distinct outflows. The HH~1002~A group is moving slowly to the right (northerly) in Figure~\ref{fig:pms_field2}, roughly orthogonal to the axis of the bipolar HH~1002~B/C outflow. It is possible that the HH~1002~A motions represent sidesplash in the extreme wings of an older, no-longer-visible, east-moving outburst, however the southern cavity edges of HH~1002~B show no such analogous motion in the opposite direction. Future proper-motion and/or radial-velocity measurements could further elucidate whether HH~1002~A represents a distinct outflow.

The optical HH~1002~B \& C features move relatively slowly in opposite directions away from the Class~I YSO SPICY~7441. A single arc of \mh\ emission near the YSO among the HH~1002~C filaments is identified as MHO~1647 in Figure~\ref{fig:MHOc16}.
Emission from the HH~1002~C bow shock in the F200W filter closely resembles \pa\ and likely traces Br$\gamma$ emission within the bandpass. The proper-motion velocities of the opposing HH~1002~B \& C features are similar in magnitude, so assuming ballistic motion at the currently measured speeds, the kinematic ages are $\sim$5300 and 830 years, respectively. 
\\

\textit{HH~1222:}
Adjacent to the HH~1002~B side of the bipolar outflow lies a series of knots labeled `jet' by \citet{smith2010}.
These knots emerge from the cloud edge with transverse speeds of $\sim$100~\kms. 
With this confirmation of its jet-like nature, the `jet' is given the designation HH~1222. 
There is no clear IR driving source for HH~1222 near the edge of the molecular cloud. 
Tracing the knot paths back to an apparent cavity at the cloud edge, we estimate kinematic timescales of 400, 550 and 670 years for the three knots we can measure since they emerged into the H~{\sc ii} region.
The direction of propagation of the HH~1222 knots is slightly tilted southward a few degrees compared to the brightest filament near the apex of HH~1002~B.
This filament contains two brighter knots and then a third fainter knot is $\sim$1\arcsec\ to the left (southward) in Figures~\ref{fig:pms_field2} and \ref{fig:MHOc16}a. Blinking the \pa\ and H$\alpha$ images hints at a slightly divergent trajectory for the fainter knot, suggesting this may be another (older) HH~1222 knot. This knot would then have a kinematic age of $\sim$3300 yrs since emerging into the H~{\sc ii} region if its average speed were between 50-100~\kms.
\\

\textit{HH~1221:}
In the very lower-right of Figure~\ref{fig:pms_field2} is an arcuate feature with a bright filament labeled HH~1221. This feature has a northerly proper motion, and is of similar scale and speed but propagates in the opposite direction of HH~1003~A. As shown in Figure~\ref{fig:MHOc9} and discussed in Section~\ref{ss:H2jets}, we argue that HH~1221 is part of a bipolar outflow with optical bow shocks HH~1003~A and HH~1221, and the collimated molecular outflow MHO~1639, emanating from the IR source J103653.8-583748.
At a projected distance of $\sim$52\arcsec\ ($\sim$0.6~pc) from the IR source suspected to drive the MHO~1639 outflow, the HH~1221 bow shock traveling at $\sim$100~\kms\ has a kinematic age of $\sim$5600~years. On the opposite side of the IR source at a projected distance of $\sim$23\arcsec\ ($\sim$0.25~pc), the bright HH~1003~A bow shock traveling at a transverse speed of $\sim$115~\kms\ has a kinematic age of $\sim$2200~years.

\subsubsection{PM Field 3}
\label{sss:pmf3}
\textit{HH~1223:}
Further to the north of the HH~1002 features, \citet{smith2010} identify shock-like features emerging from the cloud edge as the candidate outflow HH~c-1. 
Comparison with continuum-subtracted \mh\ images shows that optical emission features sit just south of the embedded MHO~1643 structures and extend outside the cloud into the H~{\sc ii} region (see Figure~\ref{fig:MHOc13}b and \ref{fig:pms_field3}). 
The morphologies of HH~c-1 and MHO~1643 suggest misaligned axes tracing separate flows. 
Proper motions demonstrate that filaments in the HH~c-1 complex are moving, indicating supersonically outflowing gas even though no collimated jet body is seen in the feature. 
With proper motions confirming the flow, this object is now designated HH~1223. 
The several arcs just east of SPICY~7440 (above in Figures~\ref{fig:MHOc13} and \ref{fig:pms_field3}) are slow-moving at $\sim$35 \kms\ (kinematic age of $\sim$1200 years) toward the cloud edge, but in the same direction as the faint structures traveling $\sim$50~\kms\ about $10\arcsec$ beyond the cloud edge in the H~{\sc ii} region (kinematic age of $\sim$4300 year).

The elongated structure recognized by \citet{smith2010} for its relatively high [S~{\sc ii}]/H$\alpha$ ratio shows little or no proper motion, and may lie at the edge of one or more intersecting outflow cavities. 
\\

\textit{HH~1218:}
The filaments at the edge of the cloud just northward show motions of $\sim$100~\kms\ in a direction consistent with their being part of the MHO~1643 outflow. Their kinematic age from the J103654.2-583626 IR source, which is visible in the \pa\ image, is approximately 1400 years.
On the opposite side of the MHO~1643 driving source about equidistant from these faster cloud edge filaments lies a faint arc of emission that forms a partial bow shock. A distinct knot in this partial H-emitting bow shock shows a transverse velocity of $\sim$70~\kms\ in a direction opposite to the cloud edge filaments, and consistent with J103654.2-583626 being the driving source of the bipolar flow. The kinematic age of the H-emitting partial bow shock is $\sim$2300 years.
When compared to the continuum-subtracted \mh\ images, it is immediately clear that this partial bow coincides with the biggest and brightest \mh\ bow shock associated with the western side of MHO~1643 (see Figures~\ref{fig:MHOc13} and \ref{fig:pms_field3}).
We therefore infer that the molecular bow shock that spatially coincides with the partial optical bow shock has the same 2300-yr kinematic age, and that the trailing molecular bow shock likely has a kinematic age about 2/3 as long, or $\sim$1500 years, similar to the faster optical filaments at the cloud edge.
\\

\textit{HH~c-5:}
Intriguingly, a small arc of emission projects just southward (left in Figure~\ref{fig:pms_field3}) of the YSO SPICY~7440, a flat spectrum source that is visible in both H$\alpha$ and \pa\ images.
Tenuous emission from the wings of the shock may connect back toward this YSO, like a small wind-blown cavity. The apex of the shock is also apparent in the F200W image, perhaps tracing shock-excited \mh, Br$\gamma$ emission, or both. 
Any counterpart in the narrowband F470N image is obscured by the airy rings of the saturated stars in the cluster. 
At first glance, this jet appears one-sided and has an estimated proper-motion velocity of $\sim$200~\kms\ ($\sim$10~pixels), the fastest of any feature. 
Additional \pa\ emission on the opposite side of the YSO may trace shock-excited emission from a counter-jet (see Figure~\ref{fig:pms_field3}), but this may be confused with \pa\ emission from the base of MHO~1643. 

\begin{figure}
	\includegraphics[width=0.5\columnwidth]{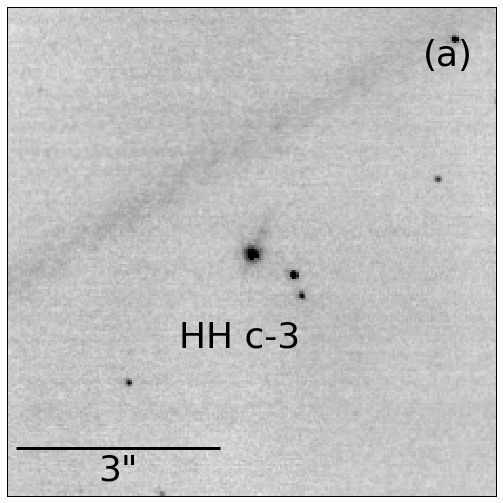}
	\includegraphics[width=0.5\columnwidth]{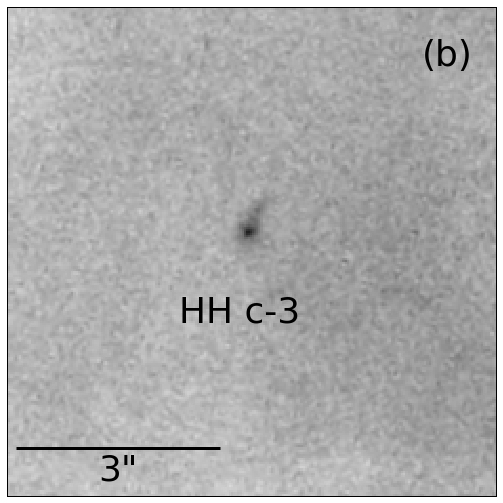}
	\includegraphics[width=0.5\columnwidth]{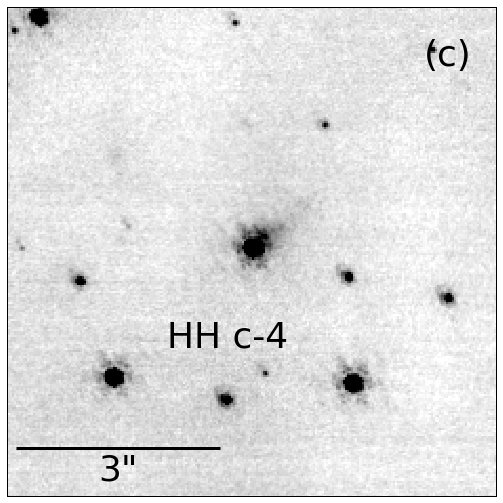}
	\includegraphics[width=0.5\columnwidth]{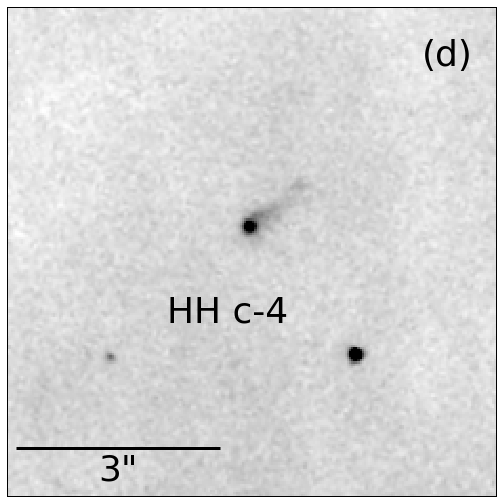}
	\includegraphics[width=0.5\columnwidth]{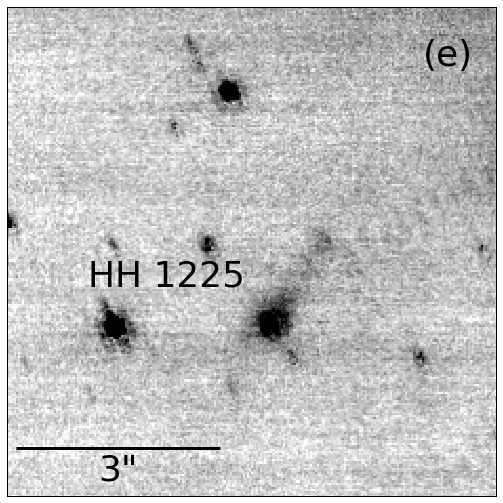}
	\includegraphics[width=0.5\columnwidth]{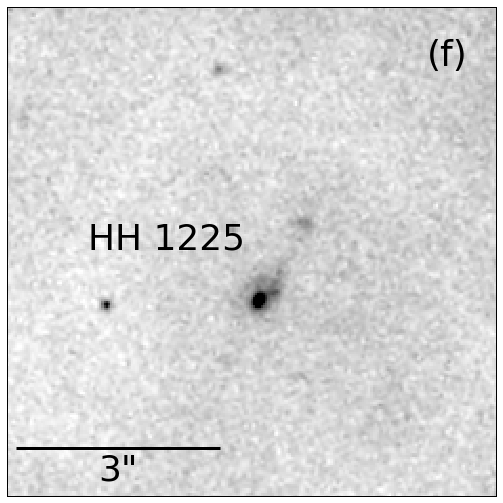}
    \caption{The candidate microjets shown in \pa\ in the left column and \ha\ in the right column. From top to bottom: HH~c-3, HH~c-4, and HH~1225. 
    }
    \label{fig:microjets}
\end{figure}

\subsubsection{Microjets}
\label{sss:microjets}
We identify two candidate and one confirmed microjet in the H$\alpha$ and \pa\ images: HH~c-3, HH~c-4, and HH~1225 (see Table~\ref{t:microjets} and Figure~\ref{fig:microjets}). 
All three appear monopolar, with collimated emission that extends $< 0.5\arcsec$ ($<1150$\,AU) from stars that are visible in the H~{\sc ii} region. 
Only candidate HH~c-4 is associated with a YSO, SPICY~7467, a Class~II source.  
All three microjet axes have similar orientations (see Figure~\ref{fig:microjets}) and a morphology that is reminiscent of the photoevaporating protoplanetary disks (proplyds) seen in Orion and other nearby H~{\sc ii} regions \citep{1994ApJ...436..194O, 2000AJ....119.2919B, 2016ApJ...826L..15K, 2021MNRAS.501.3502H}. 
However, proplyds have a cometary tail on the far side from the UV source, which would be in the opposite direction of what we observe for the ionizing sources in NGC~3324 (see Section~\ref{s:not_carina}). 

In general, emission from the candidate microjets is too faint and too smooth to reliably measure proper motions. 
Only HH~1225 has a clearly defined knot at the tip of the microjet, similar to HH~1018 \citep[see][]{smith2010,reiter2017}. 
Using this knot, we estimate a jet velocity of $\sim$100~\kms\ 
($\sim$5 pixels) along PA $\sim$ 66$^{\circ}$.

\begin{table*}
	\centering
	\caption{Previously identified and new candidate HH jets in NGC~3324.}
	\label{t:microjets}
	\begin{tabular}{lccrrl} 
		\hline
		Designation & R.A. & Dec. & YSO & Stage & Comment \\
		\hline
		HH~1002$^{\dagger}$ & 10:36:57.1 & $-$58:37:26 & J103654.0-583720 & Class~I & SPICY~7441 \\ 
		HH~1003~A$^{\dagger}$ & 10:36:53.6 & $-$58:38:09 & J103653.8-583748 & ... & Associated with HH~1221 and MHO~1639 \\ 
		HH~1003~B & 10:36:53.6 & $-$58:38:45 & J103653.3-583754 & uncertain & SPICY~7438 \\ 
		HH~1003~C$^{\dagger}$ & 10:36:54.8 & $-$58:39:09 & J103653.3-583754 & uncertain & SPICY~7438 \\ 
		%
		HH~1218 & 10:36:52.6 & $-$58:36:20 & ... & ... & Associated with the bow shock of MHO~1643 \\ 
		HH~1219 & 10:36:52.8 & $-$58:38:13 & J103652.3-583809 & Class~I & SPICY~7434; proper motions make clear that it is \textit{not} part of HH~1003~A \\ 
		HH~1220 & 10:36:53.5 & $-$58:37:60 & ... & ... & May be part of MHO~1639 \\ 
		HH~1221 & 10:36:54.6 & $-$58:36:57 & ... & ... & Part of MHO~1639 \\ 
		HH~1222$^{\dagger}$ & 10:36:54.8 & $-$58:37:18 & ... & ... & Labeled `jet' in \citet{smith2010}, near HH~1002 but kinematically distinct \\ 
		HH~1223$^{\dagger}$ & 10:36:55.9 & $-$58:36:38 & J103653.9-583629 & flat spectrum & SPICY~7440; HH~c-1 in \citet{smith2010} located south of MHO~1643 \\ 
		HH~1224$^{\dagger}$ & 10:37:01.1 & $-$58:38:37 & ... & ... & HH~c-2 in \citet{smith2010} \\ 
		HH~1225 & 10:37:01.7 & $-$58:39:31 & ... & ... & Microjet with \ha\ and \pa\ \\ 
		%
		%
		HH~c-3 & 10:37:01.5 & $-$58:37:51 & J103701.5-583751 & ... & Candidate microjet seen in H$\alpha$ and \pa\ \\ 
		HH~c-4 & 10:37:02.1 & $-$58:36:57 & J103702.1-583658 & Class~II & SPICY~7467; candidate microjet seen in H$\alpha$ and \pa\ \\ 
		HH~c-5 & 10:36:53.9 & $-$58:36:32 & J103653.9-583632 & ... & \pa\ and visible in F200W (Br$\gamma$?) \\ 
		\hline
		\multicolumn{6}{l}{$^{\dagger}$ identified in \citet{smith2010}}
	\end{tabular}
\end{table*}

%
\begin{table*}
	\centering
	\caption{Proper motions of H-emitting outflow features}
	\label{t:pms}
	\begin{tabular}{lcccl} 
		\hline
		HH          & $v_T$  & P.A.         & Kinematic & Comment \\
		Designation & (\kms)$^{\dagger}$ & ($^{\circ}$)$^*$ & Age (yrs) &  \\
		\hline
		HH~1002~A & $\sim$25 & 17 & - & Driving source uncertain; sidesplash of older outburst? \\ 
		HH~1002~B & $\sim$50 & 107 & 5300 &  \\ 
		HH~1002~C & $\sim$57 & 286 & 830 &  \\ 
		HH~1222 & $\sim$100 & 119 & 400/550/670 & 3 jet knots form a distinct outflow; may include faint knot in HH~1002~B complex \\ 
		HH~1003~A & $\sim$115 & 182 & 2,200 &  Associated with MHO~1639 and HH~1221 \\ 
		HH~1219 & $\sim$85, 100 & 139 & 770/1,100/1,650/2,000 & 4 features measured \\ 
		HH~1003~B & $\sim$115 & 165 & 2,800-3,800 & 2 shock complexes; foreground star unrelated? \\ 
		HH~1003~C & $\sim$50 & 160 & 7,000-14,000 & Younger age range assumes deceleration has occurred \\ 
		HH~1223~ridges & $\sim$35 & 110 & 1200 & Uneven expansion across filaments \\
		HH~1223~left/slow & $\sim$50 & 105 & 4300 & Diffuse features, uncertain measurements \\
		HH~1218-counterjet & $\sim$100 & 112 & 1400 & Fast filaments at cloud edge, part of MHO~1643  \\
  	    HH~1218 & $\sim$70 & 295 & 2300 & Part of MHO~1643, coincides with \mh\ bow shock \\
		HH~1224 & $\sim$140 & 152 & 6,000 & Counter flow to MHO~1642? \\
		HH~1220 & $\sim$130 & 185 & - & Small bow shock, uncertain driving source, may be associated with MHO~1639 \\ 
		HH~1221~east  & $\sim$100 & 355 & 5600 & Associated with MHO~1639 and HH~1003~A \\
		HH~1225 & $\sim$100 & 66 & 160 & Knot at tip of microjet \\
		HH~c-5 & $\sim$200 & 190 & 150 & Variable emission pattern may not be shock motion  \\
		\hline
        \multicolumn{5}{l}{$^{\dagger}$ Transverse velocity uncertainty $\approx \pm20$ \kms} \\
		\multicolumn{5}{l}{$^*$ Typical Position Angle uncertainty is $\pm5^{\circ}$}
	\end{tabular}
\end{table*}

\section{Discussion}\label{s:discussion}

Near- and mid-IR observations from \emph{JWST} penetrate the dusty walls surrounding the NGC~3324 H~{\sc ii} region, revealing more than a dozen \mh\ outflows for the first time. 
We identified \mh\ flows based on their morphology in F470N--F444W images. 
Streams of \mh\ knots and striking bow shocks trace \Njets\ distinct outflows. 
Many of the outflows are clearly bipolar (12/\Njets) 
while in other cases, only a single limb can be identified (5/\Njets). 
Sources like MHO~1648 that do not have an obvious driving source are more difficult to classify. 
Other sources, like MHO~1637 and MHO~1646, have bright emission along the probable outflow axis on the opposite side of the driving source that may trace the counterflow. 
However, these counterflows are difficult to confirm because they are located amid the complex subtraction residuals of the \emph{JWST} PSF.

Several of the outflows seen with \emph{JWST} are clustered together with flow axes that overlap in projection. 
For sources like MHO~1636, 1637, and 1638 (see Figure~\ref{fig:MHOc6}), proper motions measured with future epochs of \emph{JWST} imaging will clarify the membership of individual shock features in each of the outflows. 
Proper motion measurements will also clarify the origin of the chain of knots seen between MHO~1634 and MHO~1635 (see Figure~\ref{fig:MHOc4}). 
Finally, proper motions are one of the best ways to confirm or refute candidate driving sources listed in Table~\ref{t:jets}  \citep[e.g.,][]{reiter2017}.

In addition to the \mh\ flows, we also identify a few irradiated outflows and shock-like features in \pa. 
Many of the most prominent shocks were identified by \citet{smith2010} in their \ha\ images. 
For features detected in both \ha\ and \pa, we measure proper motions to confirm their jet-like nature. 
In this way, we identify five new HH jets and one new candidate. 
Finally, we identify three candidate microjets in the H~{\sc ii} region (see Figure~\ref{fig:microjets}). 
In retrospect, all three are also visible in the \ha\ images from \emph{HST}.

The exquisite angular resolution and IR sensitivity of \emph{JWST} allow us to directly compare the atomic and molecular outflow components, providing a more complete and, in some cases, more complex picture.  
For example, MHO~1647 consists of a bow shock propagating from the SPICY~7441~YSO into the cloud (see Figure~\ref{fig:MHOc16}). 
This \mh\ shock sits within the larger bipolar HH~1002 outflow identified by \citet{smith2010}. 
The counterflow propagating into the H~{\sc ii} region is only seen in \pa\ and \ha.

Like MHO~1647, MHO~1639 has both atomic and molecular components. 
Inside the cloud, a chain of \mh\ knots trace a collimated outflow axis that extends the width of one of the `mountains' in these `cosmic cliffs' (see Figures~\ref{fig:overview} and \ref{fig:MHOc9}). 
Two shocks in the H~{\sc ii} region seen only in \pa\ and \ha, HH~1003~A and HH~1221, lie along the same straight jet axis traced by MHO~1639. 
Proper motions of the two atomic shocks trace an axis that coincides with the jet-like \mh\ emission.  

Most of the embedded \mh\ outflows do not have associated \pa\ emission. 
Instead, \pa\ appears to trace either a cavity or a reflection nebula around some of the driving sources (e.g., MHO~1634, MHO~1639, and MHO~1643 see Figures~\ref{fig:MHOc4}, \ref{fig:MHOc9}, and \ref{fig:MHOc13}, respectively). 
In shocks where both \mh\ and \pa\ are observed, their ratio may be used to estimate the shock type \citep[as in][]{colgan2007}. 
However, only the prominent bow shocks of MHO~1643 have weak associated \pa, so a full excitation analysis is not possible with these data.

Almost all of the new candidate outflows that we report in NGC~3324 are completely embedded. 
This is in contrast to the many HH jets that reside in the main portion of the Carina star-forming complex \citep{smith2010,reiter2016,reiter2017}. 
These jets were discovered in \ha\ images and are thus overwhelmingly seen near cloud edges if not in the H~{\sc ii} region entirely. 
\citet{reiter2016} used near-IR [Fe~{\sc ii}] observations to trace the embedded portion of the Carina jets, in many cases connecting shock-like wisps seen in \ha\ outside the cloud 
to the IR-bright driving source itself.  
Numerous \mh\ flows have also been detected in the main portion of the Carina star-forming complex \citep{preibisch2011_hawki,tapia2011,har15}.

Finally, we note that the NGC~3324 region as a target for \emph{JWST} ERO observations was not chosen as a famous site for protostellar outflow activity. Based on wide-field mid-IR emission, the most active star formation in the region is further south along the  bubble rim (see Figure~\ref{fig:overviewBIG}b, and \citealt{sb07}). 
As such, these observations may represent the level of outflow activity we expect to see with \emph{JWST} observations of star-forming regions. Most of the outflows we identified in NGC~3324 are seen in \mh\ with no \ha/\pa\ counterpart, so we expect that the high angular resolution at IR wavelengths afforded by \emph{JWST} is likely to significantly increase the census of outflows in places like main Carina star-forming complex too.
Future observations of NGC~3324 with \emph{JWST} at other wavelengths and later epochs will enhance the value of this dataset by providing proper motions of embedded outflow components and a clearer identification of the outflow-driving sources.

\subsection{Outflow bending and  orientation}\label{ss:jet_PAs}

Many of the \mh\ flows in NGC~3324 have bent outflow axes. 
None trace the characteristic S-shape typical of a precessing jet \citep[e.g.,][]{raga1993,terquem1999}. 
Instead, the axes of MHO~1632, MHO~1638, MHO~1644, and MHO~1649 trace gentle C-shaped arcs while MHO~1633 has a more sharply curved J-shape. 
Overall, 5/\Njets\ (21\%) appear bent, similar to the fraction found in Cygnus~X by \citet{makin2018}. 
This is higher than the fraction of bent outflows in lower-mass regions \citep[14\%,][]{froebrich2016} which \citet{makin2018} speculate may be due to deflection by dust in the environment.

\citet{eisloffel2000} also find that bent outflows are common in their sample of parsec-scale \mh\ outflows.
They propose two mechanisms that may explain the shape of outflows where both lobes of the bipolar flow arc in the same direction. 
First, a supersonic side-wind \citep{canto1995,salas1998} may bend the outflows, as has been proposed to explain bent jets in the Orion H~{\sc ii} region \citep{bally2006}. 
This seems unlikely for the embedded jets in NGC~3324 as they are shielded by the surrounding molecular cloud. 
Second, the motion of the driving source itself may lead to an apparent bend of the outflow axis. 
In this scenario, the driving source ejects a straight bipolar outflow. 
Once launched, the outflow knots continue to travel ballistically. 
However, the star is not stationary, so the next knot ejection happens when the star has moved away from the location (and outflow axis) seen at a time $t=0$.
These subsequent bursts appear offset from the previous outflow knots, tracing out an apparently curved flow. 
\citet{reiter2020_tadpole_comp} proposed a similar scenario to explain the apparent bending of HH~900. 
Proper motions of the \mh\ outflows and their driving sources are required to test this hypothesis.

\begin{figure}
    \centering
    \includegraphics[width=\columnwidth]{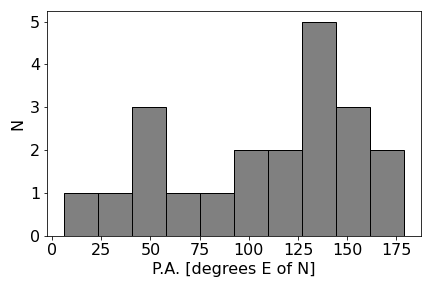}
    \caption{The distribution of P.A.s of the new \mh\ outflows. 
    }
    \label{fig:mho_pas}
\end{figure}
We measure the outflow orientation by defining an axis that connects the two most distant outflow knots identified. 
Sharply curved sources like MHO~1633 are therefore given only one position angle (P.A.).  
More precise estimates will be possible when proper motions are available to confirm the knot membership of outflows like MHO~1633. 
For most of the bent outflows in this sample, the P.A. measured this way is approximately the same as a tangent line to the outflow at the location of the driving source. 
Outflow P.A.s in degrees east of north are listed in Table~\ref{t:jets} and the distribution of outflow P.A.s is shown in Figure~\ref{fig:mho_pas}. 

The measured P.A.s are distributed over almost the full range from $0^{\circ}-180^{\circ}$. 
Two notable peaks near $50^{\circ}$ and $130^{\circ}$ are enhanced by the nearly parallel outflow pairs, MHO~1634 and MHO~1635 and MHO~1640 and MHO~1641 (discussed below). 
A Kolmogorov-Smirnov (K-S) test indicates that the distribution of P.A.s cannot be distinguished from a uniform random distribution (p-value=0.083). 
While we caution that this sample is small, multiple studies have P.A. distributions consistent with a random distribution  \citep[e.g.][]{davis2009,froebrich2016,stephens2017,baug2020} 
although preferential outflow orientations have been found with respect to local dust lanes and filaments in young high-mass regions 
\citep[e.g.,][]{davis2007,raga2010,makin2018,kong2019}.

While the population of outflows overall traces a  wide range of P.A.s, two pairs of outflows show remarkable alignment: MHO~1634 and MHO~1635 (see Figure~\ref{fig:MHOc4}) 
and 
MHO~1640 and MHO~1641 (see Figure~\ref{fig:MHOc11}). 
As projected on the plane of the sky, the outflows appear almost parallel with projected separations between the outflow axes of $\sim$10,000 -- 20,000~AU. 
The outflow origins and candidate driving sources are also located in close proximity although not perfectly aligned. 
Only $\sim$6.5\arcsec\ ($\sim$15,000~AU)
separate the MHO~1634 and MHO~1635 driving sources. 
The candidate YSOs driving MHO~1640 and MHO~1641 are further apart, with the closer of the two candidate driving sources for MHO~1641 located $\sim$13\arcsec\ ($\sim$30,000~AU) from the MHO~1640 driving source. 
For comparison, in Cygnus~X \citet{makin2018} find 10 parallel outflows with driving sources separated by $<$20\arcsec\ ($<$28,000~AU at 1.4~kpc), representing 2\% of their sample. 
The four outflows in parallel pairs represent a larger fraction of our smaller sample: 17\% or 4/\Njets.

The separations between outflow axes and their driving sources are larger than typical core sizes ($\sim$10,000~AU). 
If these neighboring stars formed from the same core, then they must have migrated outward. 
However, there are a couple of challenges to this interpretation. 
First, at smaller separations, the protostars and their disks will exert a stronger gravitational influence on each other. 
This may lead to precession causing the outflows to trace an S-shape on the sky. 
We do not observe this. Instead, the outflows in these pairs appear quite straight (especially MHO~1640 and MHO~1641, see Figure~\ref{fig:MHOc11}). 
Indeed, even in the absence of the S-shape, observations suggest that outflows from binary pairs are either randomly or preferentially anti-aligned \citep{lee2016}. 
Second, as discussed above, if the driving sources are moving, this may lead to an apparent bend in the outflow. 
None of the outflows in these parallel pairs show a strong C-shape. 
Nevertheless, driving source motion may play a role in
the MHO~1634 and MHO~1635 pair. 
The two flows have broader outflow lobes that may faintly arc toward each other. 
More distant \mh\ knots lie almost exactly between the two outflow axes (see Figure~\ref{fig:MHOc4}) making it unclear which source drives them. 
Proper motions are required to constrain the motion of the driving sources and identify which outflow likely drives [which of] the more distant blobs. 
Finally, we note that the current separation between the driving sources in both pairs is larger than the typical binary separations for all but the highest mass stars \citep[see, e.g., Figure~2 in][]{duchene2013}.

The cluster of outflows containing MHO~1643, MHO~1644, HH~1223, and HH~c-5 represents the contrasting scenario, more similar to the quadrupolar flows reported in \citet{froebrich2016,makin2018}. 
Multiple outflows emerge from the same cluster but the outflow axes are all oriented in different directions. 
MHO~1643 and MHO~1644 trace the characteristic X-shape of quadrupolar flows. 
Both are noticeably bent with more distant knots arcing further to the south (left in Figure~\ref{fig:MHOc13}). 
HH~1223 propagates along a third axis between MHO~1643 and MHO~1644 and candidate jet HH~c-5 is nearly perpendicular to the two \mh\ outflows. 
Future proper motion measurements will help clarify the complex outflow motions in this region.

MHO~1645 and MHO~1646 trace another X-shaped quadrupolar flow. 
In total, 17\% (4/\Njets) of the MHOs in NGC~3324 are in quadrupolar systems, somewhat higher than the fraction (9--10\%) reported by \citet{froebrich2016} and \citet{makin2018}.

Outside the cloud, 
HH~1003~A, B, \& C, together with HH~1220, HH~1221 and MHO~1639, may represent a parsec-scale bipolar outflow that has undergone directional changes either due to precession of the driving source or due to the external environment. 
In general, the proper-motion speeds are higher closer to the suspected driving source(s) and the furthest structures, such as HH~1003~C, are the slowest moving and largest in size. Such characteristics are common to other (more isolated) parsec-scale outflows such as HH~34 or HH~47 \citep{bally1994,stanke1999}.

Outflows in the H~{\sc ii} region may be deflected by the photoablation flows originating at the ionization front.  
Propagation through the resulting density gradient may gradually ‘bend’ the trajectory of the outflowing material away from the cloud. 
Photoevaporated material flows away from the cloud edge at the sound speed ($\sim$10--15 \kms) and imparts an additional impulse on the cross-cutting outflow features 
\citep[e.g.,][]{hester1996,bally2006,smith2010}.

However, it seems more likely that HH~1003~B \& C are a separate flow from MHO~1639 and its associated HH objects, HH~1003~A and HH~1221 (and possibly HH~1220). 
The kinematic evidence suggests that HH~1003~A, MHO~1639, and HH~1221 form a coherent bipolar outflow with little evidence of ‘bending’ or deflection (see Figure~\ref{fig:MHOc9}). Meanwhile, HH~1003~B \& C travel in the same direction and a line connecting the apices of these bow shocks does not intersect the driving source of MHO~1639; we alternatively suggest SPICY~7438 as a potential driving source (see Figure~\ref{fig:pms_field1} and Table~\ref{t:microjets}). Moreover, HH~1003~A is traveling at about the same speed as HH~1003~B, breaking with the notion of steadily decreasing proper-motion speeds with increasing distance from the source. It seems inconsistent that a sizeable deflection of HH~1003~B would not also slow it down considerably, or that HH~1003~B used to move considerably faster in the past than HH~1003~A currently is. Future proper-motion measurements and radial velocities from spectra will help clarify the trajectory and origin of all of these features.

\subsection{YSOs}\label{ss:ysos}

The majority of known \mh\ flows are driven by young embedded sources \citep[those with a flat spectrum or positive mid-IR spectral index, e.g.,][]{davis2009}. 
To identify candidate driving sources for the new \mh\ flows presented in this paper, we search for near-IR excess sources located on the outflow axis from the SPICY catalog \citep{kuhn2021} and point sources newly identified in the \emph{JWST} ERO data.

We identify seven candidate driving sources using the SPICY catalog (see Tables~\ref{t:jets} and \ref{t:microjets}). 
Of these, 
one is a flat spectrum source and 
three are Class~I sources, 
consistent with previous results that \mh\ driving sources tend to be young. 
Of the remaining three SPICY YSOs,  
two have an uncertain evolutionary classification. 
One Class~II source appears to drive one of the candidate microjets exposed in the H~{\sc ii} region. 
From the SPICY catalog, we identify \emph{Spitzer}-detected driving sources for 3/\Njets\ MHOs and 7/\Nhh\ HH candidates that trace an additional 4 distinct flows with no molecular counterparts. 
In total, 7/31 outflows in the region (\Njets\ \mh\ outflows plus 7 distinct HH flows) have a candidate driving source identified in the \emph{Spitzer} data, corresponding to a detection rate of 23\%. 
This is much lower than in the main portion of the Carina star-forming complex where roughly half of the HH jets have a \emph{Spitzer}-detected driving source
\citep{reiter2017}.

In the area imaged 
with NIRCam, there are a total of 25 candidate YSOs in the SPICY catalog. 
Of these, 12\% (3/25) are associated with an MHO and 28\% (7/25) are associated with an outflow of any kind. 
The outflow occurrence rate among \emph{Spitzer}-detected YSOs is similar to but slightly higher than the rate found by \citet{reiter2016} for HH jets in the main portion of the Carina star-forming complex (22\%).

We also compare the location of the candidate microjets presented in this study with the location of X-ray-active young stars identified by \citet{preibisch2014}. 
X-ray emission traces an active chromosphere, as is often seen from young low-mass stars that are no longer embedded, providing a complementary sample of more evolved young stars. 
None of the coutflows or microjets presented in this study have an X-ray-active young star located on or near the outflow axis.

\subsection{Comparison to other \mh\ outflows}\label{ss:mh_comp}

Near-IR \mh\ 2.12~\micron\ emission is one of the best tracers of embedded jets and outflows. 
Spectacular examples of collimated \mh\ outflows \citep[e.g.,][]{zinnecker1998} echo structures seen in the optical \citep[e.g.,][]{reipurth1989,rei97_hh111}.
Systematic surveys for \mh\ outflow emission in the Orion star-forming complex  
\citep[e.g.,][]{yu1997,yu1999,stanke1998,stanke2002,davis2009}
and other low- and high-mass star-forming regions 
(e.g., Aquila, \citealt{zhang2015_h2}; and
DR21/W75N, \citealt{smith2014}) revealed dozens of outflows from embedded young stars. 
The largest such survey to date is the UKIRT Widefield Infrared Survey for \mh\  \citep[UWISH2,][]{froebrich2011,froebrich2015}. 
UWISH2 revealed hundreds of \mh\ flows from low- and high-mass star-forming regions 
(Serpens and Aquila,  \citealt{ioannidis2012_cat,ioannidis2012_props}; 
Cassiopeia and Auriga,  \citealt{froebrich2016}; 
M17, \citealt{samal2018}; 
Cygnus~X, \citealt{makin2018}). 
\mh\ outflows from young stars are collected in the general catalogue of molecular hydrogen emission-line objects  \citep[MHOs;][]{davis2010}\footnote{\href{http://astro.kent.ac.uk/~df/MHCat/}{http://astro.kent.ac.uk/$\sim$df/MHCat/}}, a complement to the HH catalog \citep{reipurth2000_HHcat}.

\mh\ emission lines at longer wavelengths have been used to study outflows with the InfraRed Array Camera \citep[IRAC;][]{fazio2004}
on the \emph{Spitzer Space Telescope} \citep{werner2004}. 
Shock-excited \mh\ emits strongly in the 4.5~\micron\ IRAC band allowing outflows to be identified in color-color diagrams \citep{ybarra2009} or by green emission with shock- or outflow-like morphologies \citep[using the standard color mapping where 4.5~\micron\ emission is shown in green, e.g.][]{gutermuth2008,giannini2013}. 
Extended green objects, the so-called \emph{Spitzer}~EGOs, trace outflow activity from predominantly higher-mass protostars \citep{cyganowski2008,cyganowski2009,cyganowski2011}. 
Several of these EGOs were detected in \emph{Spitzer} images of the Carina Nebula \citep{smi10b}.  Most of the \mh\ 2.12~\micron\ outflows from low-mass stars in the study of \citet{giannini2013} were also detected with \emph{Spitzer}.  
Only about half of EGOs have associated \mh\ 2.12~\micron\ emission; less than one third (28\%) of those have similar morphology between the \emph{Spitzer} and \mh\ 2.12~\micron\ emission \citep{lee2012,lee2013}. 
Instead, the broadband emission more closely resembles the near-IR continuum emission suggesting that scattered light contributes to the extended emission seen in the [4.5] \emph{Spitzer} band.

The new \mh\ outflows presented here were all identified in continuum-subtracted narrowband images. 
Some knots are also visible in F200W images, likely tracing either \mh\ 2.12~\micron\ or Br$\gamma$ 2.16~\micron\ emission in the band 
(e.g., MHO~1643 and MHO~1647, see Figure~\ref{fig:MHOc13} and \ref{fig:MHOc16}, respectively). 
None of the flows reported here were previously identified as EGOs \citep[e.g., by][]{kuhn2021}. 
This is consistent with these outflows being driven by low- to intermediate-mass YSOs.

We note two bright reflection nebulae that stand out in color images (yellow in Figure~\ref{fig:overview}). 
Both are at the origin of \mh\ flows. 
The bright reflection nebula from SPICY~7423, the MHO~1634 driving source, surrounds a collimated chain of \mh\ knots that trace the outflow axis. 
A similar bowl of emission is seen at the origin of MHO~1643 and MHO~1644 (see Figure~\ref{fig:MHOc13}). 
Flux in the continuum bands is more evenly distributed through the reflection nebula compared to the continuum-subtracted \mh\ images.

No other narrowband \mh\ filters (F212N, F323N) were obtained as part of the ERO observations.
Additional observations are required for excitation analysis of the flows, as was done using  \emph{Spitzer} spectral line mapping  \citep[e.g.,][]{maret2009,nisini2010,giannini2011}.

\subsection{Outflow feedback in context}\label{ss:feedback}

Feedback from high-mass stars can trigger or accelerate star formation by compressing molecular gas  \citep[e.g.,][]{1989ApJ...346..735B, 1994A&A...289..559L, 2011ApJ...736..142B, 2011MNRAS.412.2079M, 2012MNRAS.420..562H} or it can disperse the molecular cloud and suppress star formation \citep[e.g.][]{2012MNRAS.427..625W, dale2015, grudic2021}. 
In particular, pillars and bright rimmed clouds are often suggested as sites of triggered star formation \citep[e.g.][]{2004A&A...414.1017T,2007A&A...467.1125U, 2009A&A...497..789U}. \cite{1991ApJS...77...59S} and \cite{1994A&A...289..559L} suggest an evolutionary sequence in which the irradiated clouds start with a broad morphology and become more compressed, elongated and cometary over time. 
None of the outflows in NGC~3324 are associated with the most prominent elongated  pillar in the region (the most evolved according to the above paradigm). 
This is in contrast to the many jets seen from pillar tips in the main portion of the Carina star-forming complex \citep{smith2010} which have been proposed to be the result of triggered star formation \citep[e.g.,][]{smi10b,ohl12}. 

While none of the outflows in NGC~3324 emerge from prominent pillars, the majority of the outflows are detected close to the ionization front. 
Almost all of the \mh\ outflows we report are located within $\sim$1\arcmin\ of the ionization front edge, or roughly the top third of the cloud imaged with NIRCam (see Figure~\ref{fig:overview}). 
The close association of active star formation with ionized cloud rims has been taken as evidence for triggered star formation. 
However, true triggering is difficult to prove observationally or numerically \citep[e.g.,][]{dale2015}. 
Outflows in NGC~3324 are also confined to the center of the image, with all \mh\ flows within the middle half of the north-south extent of the NIRCam mosaic. 
Future estimates of the extinction and column density in this region are required to determine whether we preferentially detect outflows in regions with lower optical depths.

Finally, with the unprecedented sensitivity of \emph{JWST}, we have detected \mh\ outflows driven by lower-mass stars despite the 2.3~kpc distance. 
Whether or not their collapse was triggered, these stars are forming from gas that was affected by feedback. 
Gas and dust temperatures are higher in high-mass star-forming regions \citep[including NGC~3324,][]{rebolledo2016} and there is 
evidence for complex organic chemistry in strongly irradiated UV environments \citep{cuadrado2017}. 
How these impact star formation is an open question, one that is getting renewed attention for its possible impact on planet formation. 
Feedback is expected to quickly expose YSOs leading to rapid disk dissipation \citep[e.g.,][]{qiao2022}. 
However, if feedback drives star-forming cores to higher densities, then they may shield their disks for a significant fraction of the planet formation timescale \citep[e.g.,][]{reiter2019_tadpole,reiter2020_tadpole,reiter2020_tadpole_comp}. 
Well-studied outflows in high-mass regions like Carina have enabled the first studies of planet-forming disks in feedback-dominated regions \citep[e.g.,][]{mesa-delgado2016,cortes-rangel2020}. 
The new population of jets in NGC~3324 provides a target list to extend these kinds of studies to a broader range of feedback conditions as well as younger and possibly lower-mass sources.

\section{Conclusions}

We report the detection of \Njets\ \mh\ outflows in NGC~3324 seen for the  first time with \emph{JWST}. 
Near-IR observations penetrate the dust wall surrounding the H~{\sc ii} region, uncovering active star formation before it is subject to the harshest feedback from the nearby O-type stars. 
We also identify \newHH\ new HH objects, complementing earlier detections made with \emph{HST}. 
In retrospect, all HH objects seen in \pa\ were also detected in earlier \ha\ imaging with \emph{HST}, allowing us to measure proper motions over a $\sim$16~yr time baseline.  

In total, we have identified 31 outflows in the  NGC~3324 region -- \Njets\ MHOs and \newHH\ HH flows. 
Our study includes \Nhh\ HH objects, including three outflows with both \mh\ and \pa\ emission (MHO~1639, MHO~1643, and MHO~1647) and 4 additional flows with no molecular counterpart. 
We identify candidate driving sources for 7/31 (23\%) of the outflows from the SPICY catalog of YSOs. 
The evolutionary classification of the candidate driving sources inferred from the shape of their IR SEDs indicates that they are young (flat spectrum or Class~I), consistent with earlier studies of \mh\ outflows. 
Using the \emph{JWST} ERO data, we identify an additional 17 point sources on or near outflow axes as candidate driving protostars.
A more detailed analysis of the physical properties of these sources is left for future work. 

The comprehensive view of the atomic and molecular components of these outflows is  only possible with \emph{JWST}.  
With the unprecedented sensitivity and angular resolution, we can compare the different components of individual outflows on the same spatial scales. 
These outflows represent a population of actively accreting stars located just outside the H~{\sc ii} region boundary, still embedded in their natal cloud sampling the moments  before they are subject to direct irradiation by the nearby high-mass stars.
This geometry has often  been taken as evidence for triggered star formation,  and the interpretation is tempting as we find few \mh\ flows deep in the cloud. 
However, we caution that other possibilities like variations in the extinction and optical depth have not been ruled out. 

Whether or not feedback triggered the collapse of the driving YSOs, these sources are forming out of gas that has been strongly affected by feedback. 
Understanding the impact of environmental feedback is currently a hot topic, especially regarding its impact on planet formation.  
These outflow-driving sources represent a well-localized set of targets for future study.  

Finally, these ERO observations represent the tip of the iceberg of what is possible for star-formation studies with \emph{JWST}. 
For NGC~3324, these images provide a first epoch for future proper motion studies that will clarify the association of individual  \mh\  knots with specific outflows and confirm or refute the candidate driving sources we identify. 
Additional observations in other filters may identify driving sources and constrain outflows excitation and mass-loss rates.
As a region with only modest star formation, NGC~3324 indicates how rich observations with \emph{JWST} stand to be for uncovering low-mass star formation in high-mass regions.

\section*{Acknowledgements}

We would like to thank John Bally for a prompt and helpful referee report. 
We would also like to thank Dirk Froebrich and Bo Reipurth for quickly providing MHO and HH numbers for this paper and their thoughtful comments on the work. 
We would also like to thank Adam Ginsburg for thoughtful discussions. 
This work is based on observations made with the NASA/ESA/CSA James Webb Space Telescope. The data were obtained from the Mikulski Archive for Space Telescopes at the Space Telescope Science Institute, which is operated by the Association of Universities for Research in Astronomy, Inc., under NASA contract NAS 5-03127 for JWST. These observations are associated with program \#2731.
This research is based in part on observations made with the NASA/ESA Hubble Space Telescope obtained from the Space Telescope Science Institute, which is operated by the Association of Universities for Research in Astronomy, Inc., under NASA contract NAS 5–26555. These observations are associated with program 10475. 
The MHO catalogue is hosted by the University of Kent. 
This research has made use of the SIMBAD database, operated at CDS, Strasbourg, France and SAOImage DS9 \citep{joye2003}. 
TJH is funded by a Royal Society Dorothy Hodgkin Fellowship. 

\section*{Data Availability}

The data used in this work are available to the public for immediate download from the MAST archive\footnote{\href{https://archive.stsci.edu/missions-and-data/jwst}{https://archive.stsci.edu/missions-and-data/jwst}}.



\bibliographystyle{mnras}
\bibliography{bibliography} 




\appendix

\section{Candidate galaxies}
\begin{figure}
	\includegraphics[width=0.240\textwidth]{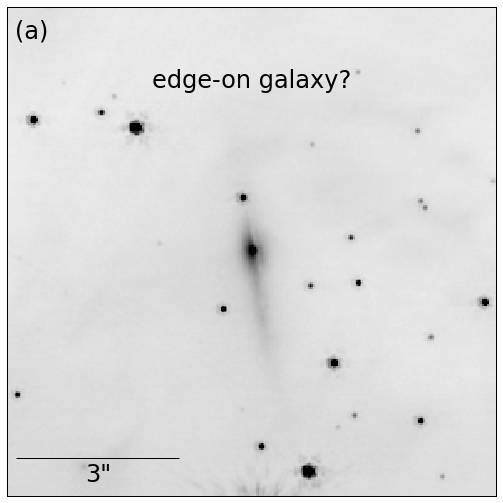}
	\includegraphics[width=0.2325\textwidth,trim=5mm 0mm 0mm 0mm]{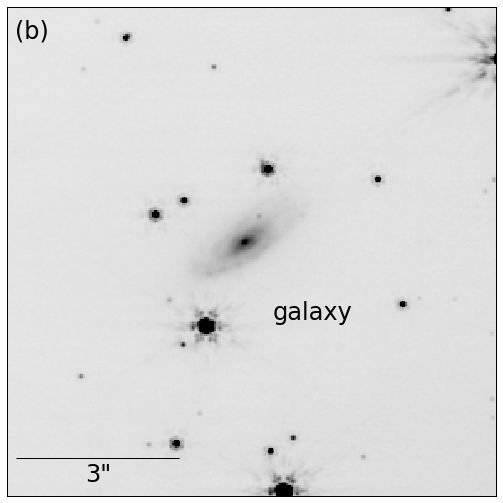}
	\includegraphics[width=0.240\textwidth]{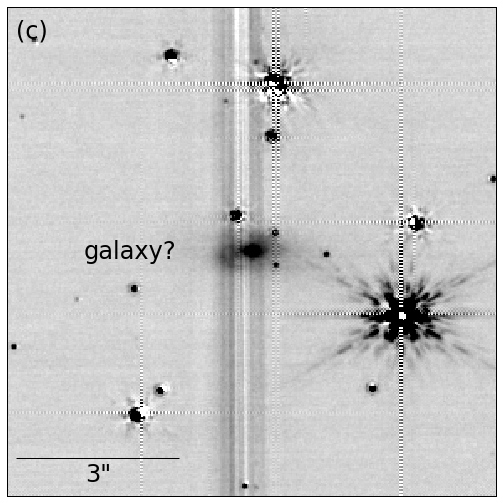}
	\includegraphics[width=0.240\textwidth]{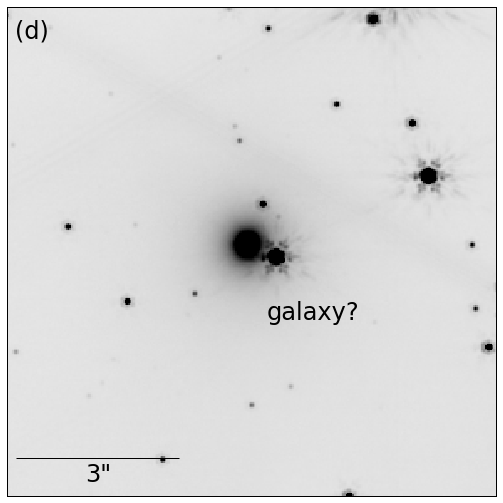}
    \caption{Candidate background galaxies shown in the F200W filter except for panel
    \textbf{(c)} which shows F200W--F090W to suppress the diffraction spike that runs through the source. 
    }
    \label{fig:edge-on}
\end{figure}
We identify four candidate background galaxies in the \emph{JWST} ERO data on NGC~3324. 
Candidates are shown in Figure~\ref{fig:edge-on}. 

An elongated object with coordinates R.A.=10:36:53.7 and Dec=$-$58:35:47 
is visible in multiple filters (F187N, F200W, F335M, F444W, F470N, F1280W, and F1800W). 
The morphology is reminiscent of an edge-on spiral galaxy (see Figure~\ref{fig:edge-on}a) but the elongated emission is asymmetrical suggesting an off-center bulge. 
Structure in the F200W image hints at a dust lane bisecting the disk-like emission but this feature is not seen at any other wavelength. 
The nearest object in the Simbad database \citep{wenger2000} is $>30\arcsec$ away from the coordinates of this object.

A second object at R.A.=10:36:59.5 and Dec=$-$58:39:01 looks like a spectacular example of a spiral galaxy (see Figure~\ref{fig:edge-on}b). 
The closest Simbad object is located $>20\arcsec$ away from these coordinates. 

A third candidate galaxy is seen at R.A.=10:37:07.2 and Dec=$-$58:35:36 (see Figure~\ref{fig:edge-on}c). 
At first glance, the object looks like a bright point source in the middle of a diffraction spike from a nearby bright star. 
However a disk-like structure can be seen in a F200W--F090W image (the source is not visible in the F090W filter so we use this to suppress the diffraction spike). 
Like the other candidate galaxies, the nearest Simbad source is offset nearly 20\arcsec\ from the position of this object. 

A candidate elliptical galaxy at R.A.=10:36:55.2 and 
Dec=$-$58:38:09 is noted in Figure~\ref{fig:pms_field1} and shown in Figure~\ref{fig:edge-on}d. The source is round and appears resolved compared to the somewhat fainter star projected next to it in the \emph{JWST} NIRCam F187N image, but is not visible in the \emph{HST} F658N image. 
The object is 0.29\arcsec\ from an RR Lyrae candidate Gaia DR3 5350681737823554176 which is likely the neighboring point source.


\bsp	
\label{lastpage}
\end{document}